\documentclass[preprint]{aastex}
\slugcomment{Accepted by the \emph{Astrophysical Journal}}
\newcommand{\MAP}{{\sl WMAP}}

\shorttitle{TESTS OF GAUSSIANITY}
\shortauthors{KOMATSU ET AL.}
\begin{document}
\title{
 FIRST YEAR {\sl WILKINSON MICROWAVE ANISOTROPY PROBE} ({\MAP}) OBSERVATIONS: 
 TESTS OF GAUSSIANITY 
}

\author{E. Komatsu \altaffilmark{2}, 
A. Kogut \altaffilmark{3}, 
M. R. Nolta \altaffilmark{4},
C. L. Bennett \altaffilmark{3}, 
M. Halpern \altaffilmark{5},
G. Hinshaw \altaffilmark{3}, 
N. Jarosik \altaffilmark{4},
M. Limon \altaffilmark{3,6}, 
S. S. Meyer \altaffilmark{7},
L. Page \altaffilmark{4},
D. N. Spergel \altaffilmark{2},
G. S. Tucker \altaffilmark{3,6,8},
L. Verde \altaffilmark{2,9},
E. Wollack \altaffilmark{3},
E. L. Wright \altaffilmark{10}
}

\altaffiltext{1}{
{\MAP}  is the result of a partnership between Princeton 
 University and NASA's Goddard Space Flight Center. Scientific 
 guidance is provided by the {\MAP}  Science Team.
}
\altaffiltext{2}{Dept of Astrophysical Sciences, Princeton University, Princeton, NJ 08544}
\altaffiltext{3}{Code 685, Goddard Space Flight Center, Greenbelt, MD 20771}
\altaffiltext{4}{Dept. of Physics, Jadwin Hall, Princeton, NJ 08544}
\altaffiltext{5}{Dept. of Physics and Astronomy, University of British Columbia, Vancouver, BC  Canada V6T 1Z1}
\altaffiltext{6}{National Research Council (NRC) Fellow}
\altaffiltext{7}{Depts. of Astrophysics and Physics, EFI and CfCP, University of Chicago, Chicago, IL 60637}
\altaffiltext{8}{Dept. of Physics, Brown University, Providence, RI 02912}
\altaffiltext{9}{Chandra Fellow}
\altaffiltext{10}{UCLA Astronomy, PO Box 951562, Los Angeles, CA 90095-1562}

\email{komatsu@astro.princeton.edu}

\begin{abstract}
 We present limits to the amplitude of non-Gaussian primordial
 fluctuations in the {\MAP} 1-year cosmic microwave background 
 sky maps. A non-linear coupling parameter, $f_{\rm NL}$, 
 characterizes the amplitude of a quadratic term in the primordial 
 potential. We use two statistics: one is a cubic statistic which 
 measures phase correlations of temperature fluctuations after 
 combining all configurations of the angular bispectrum. The 
 other uses the Minkowski functionals to measure the morphology 
 of the sky maps. Both methods find the {\MAP} data consistent 
 with Gaussian primordial fluctuations and establish limits, 
 $-58<f_{\rm NL}<134$, at 95\% confidence. There is no significant 
 frequency or scale dependence of $f_{\rm NL}$. 
 The {\MAP} limit is 30 times better than {\sl COBE}, and 
 validates that the power spectrum can fully characterize 
 statistical properties of CMB anisotropy in the {\MAP} data 
 to high degree of accuracy. Our results also validate the use 
 of a Gaussian theory for predicting the abundance of clusters 
 in the local universe. We detect a point-source contribution to 
 the bispectrum at 41~GHz, 
 $b_{\rm src}=(9.5\pm 4.4)\times 10^{-5}~\mu{\rm K}^3~{\rm sr^2}$,
 which gives a power spectrum from point sources of
 $c_{\rm src}= (15\pm 6)\times 10^{-3}~\mu{\rm K}^2~{\rm sr}$
 in thermodynamic temperature units. 
 This value agrees well with independent estimates of source 
 number counts and the power spectrum at 41~GHz, indicating that 
 $b_{\rm src}$ directly measures residual source contributions.
\end{abstract}
\section{INTRODUCTION}\label{sec:intro}

The Gaussianity of the primordial fluctuations is a key assumption of 
modern cosmology, motivated by simple models of inflation.
Statistical properties of the primordial fluctuations are 
closely related to those of the cosmic microwave background (CMB) 
radiation anisotropy; thus, a measurement of non-Gaussianity of the CMB 
is a direct test of the inflation paradigm.
If CMB anisotropy is Gaussian, then the angular power spectrum fully
specifies the statistical properties.
Recently, \cite{acquaviva/etal:2002} and \cite{maldacena:2002} have 
calculated second-order perturbations during inflation
to show that simple models based upon a slowly-rolling
scalar field cannot generate detectable non-Gaussianity.
Their conclusions are consistent with previous work 
\citep{salopek/bond:1990,salopek/bond:1991,falk/rangarajan/srendnicki:1993,gangui/etal:1994}.
Inflation models that have significant non-Gaussianity 
may have some complexity such as non-Gaussian isocurvature fluctuations
\citep{linde/mukhanov:1997,peebles:1997,bucher/zhu:1997}, 
a scalar-field potential with features 
\citep{kofman/etal:1991,wang/kamionkowski:2000},
or ``curvatons'' \citep{lyth/wands:2002,lyth/ungarelli/wands:2002}.
Detection or nondetection of non-Gaussianity thus sheds light on 
physics of the early universe.

Many authors have tested Gaussianity of CMB anisotropy 
on large angular scales ($\sim 7^\circ$) 
\citep{kogut/etal:1996b,heavens:1998,schmalzing/gorski:1998,ferreira/magueijo/gorski:1998,pando/valls-gabaud/fang:1998,bromley/tegmark:1999,banday/zaroubi/gorski:2000,contaldi/etal:2000,mukherjee/hobson/lasenby:2000,magueijo:2000,novikov/schmalzing/mukhanov:2000,sandvik/magueijo:2001,barreiro/etal:2000,phillips/kogut:2001,komatsu/etal:2002,komatsu:2001,kunz/etal:2001,aghanim/forni/bouchet:2001,cayon/etal:2002}, 
on intermediate scales ($\sim 1^\circ$) 
\citep{park/etal:2001,shandarin/etal:2002}, and on small scales ($\sim 10'$) 
\citep{wu/etal:2001b,santos/etal:2002b,polenta/etal:2002}. 
So far there is no evidence for significant cosmological non-Gaussianity.

Most of the previous work only tested the consistency between the CMB data 
and simulated Gaussian realizations without having physically motivated 
non-Gaussian models. 
They did not, therefore, consider quantitative constraints on 
the amplitude of possible non-Gaussian signals allowed by the data.
On the other hand, \cite{komatsu/etal:2002}, \cite{santos/etal:2002b}, and
\cite{cayon/etal:2002}
derived constraints on a parameter characterizing the amplitude of 
primordial non-Gaussianity inspired by inflation models.
The former and the latter approaches are conceptually different;
the former does not address {\it how Gaussian} the CMB data are
or the physical implication of the results.
In this paper, we adopt the latter approach, and constrain
the amplitude of primordial non-Gaussianity in the {\MAP} 
1-year sky maps.

Some previous work all had roughly similar sensitivity to non-Gaussian CMB 
anisotropy at different angular scales, because the number of 
independent pixels in the maps are similar, i.e., $\simeq 4000-6000$ for 
{\sl COBE} \citep{bennett/etal:1996}, 
{\sl QMASK} \citep{xu/tegmark/oliveira:2002}, 
and {\sl MAXIMA} \citep{hanany/etal:2000} sky maps. 
\cite{polenta/etal:2002} used about $4\times 10^4$ 
pixels from the {\sl BOOMERanG} map \citep{debernardis/etal:2000}, 
but found no evidence for non-Gaussianity.
The {\MAP} provides about $2.4\times 10^6$ pixels (outside the 
{\it Kp0} cut) uncontaminated by the Galactic emission 
\citep{bennett/etal:2003c}, 
achieving more than one order of magnitude improvement in
sensitivity to non-Gaussian CMB anisotropy.

This paper is organized as follows.
In \S~\ref{sec:primordial}, we describe our methods for
measuring the primordial non-Gaussianity using the cubic (bispectrum)
statistics and the Minkowski functionals, and present the results of 
the measurements of the {\MAP} 1-year sky maps. 
Implications of the results for inflation models and the high-redshift 
cluster abundance are then presented.
In \S~\ref{sec:pointsources}, we apply the bispectrum to individual 
frequency bands to estimate the point-source contribution to the angular 
power spectrum. The results from the {\MAP} data are then presented,
and also comparison among different methods. 
In \S~\ref{sec:conclusion}, we present summary of our results.
In Appendix~\ref{app:simulations}, we test our cubic statistics for the
primordial non-Gaussianity using non-Gaussian CMB sky maps directly
simulated from primordial fluctuations.
In Appendix~\ref{app:ptsrc}, we test our cubic statistic for the 
point sources using simulated point-source maps.
In Appendix~\ref{app:step}, we calculate the CMB angular bispectrum 
generated from features in a scalar-field potential.

\section{LIMITS ON PRIMORDIAL NON-GAUSSIANITY}\label{sec:primordial}

\subsection{The {\MAP} 1-year Sky Maps}\label{sec:map_data}

We use a noise-weighted sum of the Q1, Q2, V1, V2, W1, W2, W3, and W4 maps.
The maps are created in the HEALPix format with $nside=512$
\citep{gorski/hivon/wandelt:1998},
having the total number of pixels of $12\times nside^2=3,145,728$.
We do not smooth the maps to a common resolution before forming the
co-added sum.
This preserves the independence of noise in neighboring pixels,
at the cost of complicating the effective window function for the sky signal.
We assess the results by comparing the {\MAP} data
to Gaussian simulations processed in identical fashion.
Each CMB realization draws a sample from the $\Lambda$CDM cosmology
with the power-law primordial power spectrum fit to the {\MAP} data 
\citep{hinshaw/etal:2003,spergel/etal:2003}.
The cosmological parameters are in Table~1 of 
\cite{spergel/etal:2003} (we use the best-fit ``{\MAP} data only'' parameters).
We copy the CMB realization and smooth each copy
with the {\MAP} beam window functions of the Q1, Q2, V1, V2, 
W1, W2, W3, and W4 \citep{page/etal:2003b}.
We then add independent noise realizations to each simulated map, 
and co-add weighted by $N_{\rm obs}/\sigma_0^2$, where
the effective number of observations $N_{\rm obs}$ varies across the sky.
The values of the noise variance per $N_{\rm obs}$, $\sigma_0^2$,
are tabulated in Table~1 of \citep{bennett/etal:2003b}.

We use the conservative $Kp0$ mask to cut the Galactic
plane and known point sources, as described in
\cite{bennett/etal:2003c}, retaining 76.8\% of the sky 
(2,414,705 pixels) for the subsequent analysis. 
In total 700 sources are masked on the 85\% of the sky outside the 
Galactic plane in all bands; thus, the number density of masked sources is 
65.5~${\rm sr}^{-1}$.
The Galactic emission outside the mask has visible 
effects on the angular power spectrum \citep{hinshaw/etal:2003}.
Since the Galactic emission is highly non-Gaussian, we need to reduce
its contribution to our estimators of primordial non-Gaussianity.
Without foreground correction, both the bispectrum and the Minkowski 
functionals find strong non-Gaussian signals. 
We thus use the forground template correction described
in section~6 of \cite{bennett/etal:2003} to reduce foreground 
emission to negligible levels in Q, V, and W bands.
The method is termed as an "alternative fitting method", which
uses only the Q, V, and W band data. The dust component is 
separately fitted to each band without assuming spectrum
of the dust emission (3 parameters).
We assume that the free-free emission has $\nu^{-2.15}$ spectrum,
and the synchrotron has $\nu^{-2.7}$ spectrum.
The amplitude of each component in Q band is then fitted across
three bands (2 parameters).
 
\subsection{Methodology}

\subsubsection{Model for Primordial Non-Gaussianity}

We measure the amplitude of non-Gaussianity in primordial fluctuations 
parametrized by a non-linear coupling parameter, $f_{\rm NL}$
\citep{komatsu/spergel:2001}.
This parameter determines the amplitude of a quadratic term added to the 
Bardeen curvature perturbations $\Phi$ 
($\Phi_{\rm H}$ in \cite{bardeen:1980}), as
\begin{equation}
 \label{eq:phi}
 \Phi({\mathbf x})= \Phi_{\rm L}({\mathbf x}) 
 + f_{\rm NL}\left[ \Phi_{\rm L}^2({\mathbf x}) 
              - \left<\Phi_{\rm L}^2({\mathbf x})\right> \right],
\end{equation}
where $\Phi_{\rm L}$ are Gaussian linear perturbations with
zero mean.
Although the form in equation~(\ref{eq:phi}) is inspired by simple 
inflation models, the exact predictions from those inflation models are 
irrelevant
to our analysis here because the predicted amplitude of $f_{\rm NL}$ is 
much smaller than our sensitivity; however, this parameterization is 
useful to find {\it quantitative} constraints on the amount of 
non-Gaussianity allowed by the CMB data.
Equation~(\ref{eq:phi}) is general in that
$f_{\rm NL}$ parameterizes the leading-order non-linear corrections to $\Phi$. 
We discuss the possible scale-dependence in Appendix~\ref{app:step}.

Angular bispectrum analyses found 
$\left|f_{\rm NL}\right|<1500$ (68\%) from the {\sl COBE} DMR
53$+$90~GHz coadded map \citep{komatsu/etal:2002} and 
$\left|f_{\rm NL}\right|<950$ (68\%) from the {\sl MAXIMA} sky map 
\citep{santos/etal:2002b}.
The skewness measured from the DMR map smoothed with filters, called
the Spherical Mexican Hat wavelets, found 
$\left|f_{\rm NL}\right|<1100$ (68\%) \citep{cayon/etal:2002}, although
they neglected the integrated Sachs--Wolfe effect in the
analysis, and therefore underestimated the cosmic variance of $f_{\rm NL}$.
{\sl BOOMERanG} did not measure $f_{\rm NL}$ in their analysis
of non-Gaussianity \citep{polenta/etal:2002}.
The r.m.s. amplitude of $\Phi$ is given by
$\left<\Phi^2\right>^{1/2}\simeq \left<\Phi^2_{\rm L}\right>^{1/2}
\left(1+f_{\rm NL}^2\left<\Phi^2_{\rm L}\right>\right)$.
Since $\left<\Phi^2\right>^{1/2}$ measured on the {\sl COBE} scales
through the Sachs--Wolfe effect is 
$\left<\Phi^2\right>^{1/2}=3\left<\Delta T^2\right>^{1/2}/T\simeq 
3.3\times 10^{-5}$ \citep{bennett/etal:1996}, one obtains
$f_{\rm NL}^2\left<\Phi^2_{\rm L}\right><2.5\times 10^{-3}$ from
the {\sl COBE} 68\% constraints; thus, we already know that the contribution 
from the non-linear term to the r.m.s. amplitude is smaller than 0.25\%,
and that to the power spectrum is smaller than 0.5\%.
This amplitude is comparable to limits on systematic errors of the 
{\MAP} power spectrum \citep{hinshaw/etal:2003b}, and needs to be 
constrained better in order to verify the analysis of the power spectrum.

\subsubsection{Method 1: The Angular Bispectrum}

Our first method for measuring $f_{\rm NL}$ is a ``cubic statistic'' which
combines nearly optimally all configurations of the angular bispectrum
of the primordial non-Gaussianity \citep{komatsu/spergel/wandelt:2003}.
The bispectrum measures phase correlations of field fluctuations.
We compute the spherical harmonic coefficients $a_{lm}$ of temperature
fluctuations from
\begin{equation}
 \label{eq:alm}
  a_{lm}= \int d^2\hat{\mathbf n} M(\hat{\mathbf n}) 
  \frac{\Delta T(\hat{\mathbf n})}{T_0} Y_{lm}^*(\hat{\mathbf n}),
\end{equation}
where $M(\hat{\mathbf n})$ is a pixel-weighting function.
Here, $M(\hat{\mathbf n})$ is the $Kp0$ sky cut where 
$M(\hat{\mathbf n})$ takes 0 in the cut region and 1 otherwise.
We filter the measured $a_{lm}$ in $l$-space and 
transform it back to compute two new maps, 
$A(r,\hat{\mathbf n})$ and $B(r,\hat{\mathbf n})$, given by 
\begin{eqnarray}
 \label{eq:filter1}
 A(r,\hat{\mathbf n}) &\equiv& \sum_{l=2}^{l_{\rm max}}\sum_{m=-l}^l 
  \frac{\alpha_l(r)b_l}{\tilde{C}_l}a_{lm}Y_{lm}(\hat{\mathbf n}),\\
 \label{eq:filter2}
 B(r,\hat{\mathbf n}) &\equiv& \sum_{l=2}^{l_{\rm max}}\sum_{m=-l}^l 
  \frac{\beta_l(r)b_l}{\tilde{C}_l}a_{lm}Y_{lm}(\hat{\mathbf n}).
\end{eqnarray}
Here $\tilde{C}_l\equiv C_l b_l^2 + N$, where $C_l$ is the CMB anisotropy, 
$N$ the noise bias, and $b_l$ the beam window function describing the 
combined smoothing effects of the beam \citep{page/etal:2003b} and 
the finite pixel size.
The functions $\alpha_l(r)$ and $\beta_l(r)$ are defined by
\begin{eqnarray}
 \label{eq:alpha_l}
  \alpha_l(r) &\equiv&
  \frac{2}{\pi}\int k^2 dk g_{{\rm T}l}(k) j_l(k r),\\
 \label{eq:beta_l}
  \beta_l(r) &\equiv& 
  \frac{2}{\pi}\int k^2 dk P(k) g_{{\rm T}l}(k) j_l(k r),
\end{eqnarray}
where $r$ is the comoving distance.
These two functions constitute the primordial angular bispectrum
and correspond to $\alpha_l(r) = f_{\rm NL}^{-1}b_l^{\rm NL}(r)$ and 
$\beta_l(r) = b_l^{\rm L}(r)$ in the notation of \cite{komatsu/spergel:2001}.
We compute the radiation transfer function $g_{{\rm T}l}(k)$ with
a code based upon {\sf CMBFAST} \citep{seljak/zaldarriaga:1996} 
for the best-fit cosmological model of the {\MAP} 1-year data
\citep{spergel/etal:2003}.
We also use the best-fit primordial power spectrum of $\Phi$, $P(k)$.
We then compute the cubic statistic for the primordial non-Gaussianity, 
$S_{\rm prim}$, by integrating the two filtered maps over $r$ as 
\citep{komatsu/spergel/wandelt:2003}
\begin{equation}
 \label{eq:skewness}
  {\cal S}_{\rm prim} = m_3^{-1} \int 4\pi r^2 dr 
  \int \frac{d^2\hat{\mathbf n}}{4\pi}
  A(r,\hat{\mathbf n}) B^2(r,\hat{\mathbf n}),
\end{equation}
where the angular average is done on the full sky regardless 
of sky cut, and 
$m_3=(4\pi)^{-1}\int d^2\hat{\mathbf n}M^3(\hat{\mathbf n})$ is the 
third-order moment of the pixel-weighting function.
When the weight is only from a sky cut, as is the case here, we have 
$m_3=f_{\rm sky}$, i.e., $m_3$ is the fraction of the sky covered by 
observations \citep{komatsu/etal:2002}.
\cite{komatsu/spergel/wandelt:2003} show that $B$ is a Wiener-filtered
map of the underlying primordial fluctuations, $\Phi$.
The other map $A$ combines the bispectrum configurations that
are sensitive to non-linearity of the form in equation~(\ref{eq:phi}).
Thus, ${\cal S}_{\rm prim}$ is optimized for measuring 
the skewness of $\Phi$ and picking out the quadratic term in
equation~(\ref{eq:phi}).

Finally, the non-linear coupling parameter $f_{\rm NL}$ is given by
\begin{equation}
 \label{eq:f_NL}
  f_{\rm NL}
  \simeq 
  \left[
   \sum_{l_1\le l_2\le l_3}^{l_{\rm max}}
  \frac{({\cal B}_{l_1l_2l_3}^{\rm prim})^2}
  {{\cal C}_{l_1}{\cal C}_{l_2}{\cal C}_{l_3}}\right]^{-1}
  {\cal S}_{\rm prim},
\end{equation}
where ${\cal B}^{\rm prim}_{l_1l_2l_3}$ is the primordial bispectrum
\citep{komatsu/spergel:2001}
multiplied by $b_{l_1}b_{l_2}b_{l_3}$ and computed for $f_{\rm NL}=1$ and 
the best-fit cosmological model.
This equation is used to measure $f_{\rm NL}$ as a function of the 
maximum multipole $l_{\rm max}$.
The statistic ${\cal S}_{\rm prim}$ takes only $N^{3/2}$ operations to 
compute without loss 
of sensitivity whereas the full bispectrum analysis takes $N^{5/2}$ operations.
It takes about 4 minutes on 16 processors of an SGI Origin 300 
to compute $f_{\rm NL}$ from a sky map at the highest resolution level,
$nside=512$.
We measure $f_{\rm NL}$ as a function of $l_{\rm max}$.
Since there is little CMB signal compared with instrumental
noise at $l>512$, we shall use $l_{\rm max}=512$ at most; thus,
$nside=256$ is sufficient, speeding up evaluations of $f_{\rm NL}$ by a 
factor of 8 as the computational time scales as $(nside)^3$.
The computation takes only 30 seconds at $nside=256$.
Note that since we are eventually fitting for two parameters, $f_{\rm NL}$ and 
$b_{\rm src}$ (see sec.~\ref{sec:pointsources}), we include covariance 
between these two parameters in the analysis.
The covariance is, however, small (see Figure~\ref{fig:sim_hist} in 
Appendix~\ref{app:simulations}).

While we use uniform weighting for $M(\hat{\mathbf n})$, we could 
instead weight by the inverse noise variance per pixel, 
$M(\hat{\mathbf n})=N^{-1}(\hat{\mathbf n})$; however, 
this weighting scheme is sub-optimal at low $l$ where 
the CMB anisotropy dominates over noise so that the uniform weighing is 
more appropriate.
For measuring $b_{\rm src}$, on the other hand, 
we shall use a slightly modified version of the $N^{-1}$ weighting, 
as $b_{\rm src}$ comes mainly from small angular scales where instrumental 
noise dominates. 

\subsubsection{Method 2: The Minkowski Functionals}

Topology offers another test for non-Gaussian features in the maps,
measuring morphological structures of fluctuation fields.
The Minkowski functionals \citep{minkowski:1903,gott/etal:1990,schmalzing/gorski:1998}
describe the properties of regions spatially bounded by a set of contours.
The contours may be specified in terms of fixed temperature thresholds,
$\nu = \Delta T / \sigma$, where $\sigma$ is the standard deviation of the map,
or in terms of the area.
Parameterization of contours by threshold is computationally simpler,
while parameterization by area reduces correlations between the Minkowski 
functionals \citep{shandarin/etal:2002}.
We use a joint analysis of the three Minkowski functionals
(area $A(\nu)$, contour length $C(\nu)$, and genus $G(\nu)$)
explicitly including their covariance;
consequently, we work in the simpler threshold parameterization.

The Minkowski functionals are additive for disjoint regions on the sky
and are invariant under coordinate transformation and rotation.
We approximate each Minkowski functional using the set of equal-area pixels
hotter or colder than a set of fixed temperature thresholds.
The fractional area
\begin{equation}
A(\nu) = \frac{1}{A} \sum_i a_i = \frac{N_\nu}{N_{\rm cut}}
\label{area_def}
\end{equation}
is thus the number of enclosed pixels, $N_\nu$, divided by the total number 
of pixels on the cut sky, $N_{\rm cut}$.
Here $a_i$ is the area of an individual spot, and $A$ is the total area
of the pixels outside the cut. 
The contour length
\begin{equation}
C(\nu) = \frac{1}{4A} \sum_i P_i
\label{contour_def}
\end{equation}
is the total perimeter length of the enclosed regions $P_i$, while the genus
\begin{equation}
G(\nu) = \frac{1}{2 \pi A} (N_{\rm hot} - N_{\rm cold})
\label{genus_def}
\end{equation}
is the number of hot spots, $N_{\rm hot}$, minus the number of cold spots, 
$N_{\rm cold}$.
We calibrate finite pixelization effects by comparing the Minkowski 
functionals for the {\MAP} data to Monte Carlo simulations.

The {\MAP} data are a superposition of sky signal and instrument noise,
each with a different morphology.
The Minkowski functionals transform monotonically
(although not linearly) between the limiting cases of a sky signal with 
no noise and a noise map with no sky signal.
Unlike spatial analyses such as Fourier decomposition,
different regions of the sky cannot be weighted by the signal-to-noise ratio,
nor does the noise ``average down'' over many pixels.
The choice of map pixelization becomes a tradeoff
between resolution (favoring smaller pixels)
versus signal-to-noise ratio (favoring larger pixels).
We compute the Minkowski functionals at  $nside=16$ through 256
(3072 to 786,432 pixels on the full sky).  
We use the {\MAP} {\it Kp0} sky cut to reject pixels near the 
Galactic plane or contaminated by known sources.
The cut sky has 1433 pixels at resolution $nside=16$
and 666,261 pixels at $nside=256$.

We compute the Minkowski functionals at 15 thresholds 
from $-3.5\sigma$ to $+3.5\sigma$, and compare each functional to the 
simulations using a goodness-of-fit statistic,
\begin{equation}
\chi^2 = \sum_{\nu_1\nu_2}
~ [ F^i_{\rm WMAP} - \langle F^i_{\rm sim} \rangle ]_{\nu_1}
~ \Sigma^{-1}_{\nu_1\nu_2}
~ [ F^i_{\rm WMAP} - \langle F^i_{\rm sim} \rangle ]_{\nu_2},
\label{chi_def}
\end{equation}
where $F^i_{\rm WMAP}$ is a Minkowski functional from the {\MAP} data
(the index $i$ denotes a kind of functional),
$\langle F^i_{\rm sim} \rangle$ is the mean from the Monte Carlo simulations,
and $\Sigma_{\nu_1\nu_2}$ is the bin-to-bin covariance matrix from the
simulations.

\subsection{Monte Carlo Simulations}

Monte Carlo simulations are used to estimate the statistical
significance of the non-Gaussian signals.
One kind of simulation generates Gaussian random 
realizations of CMB sky maps for the angular power spectrum, window functions, 
and noise properties of the {\MAP} 1-year data.
This simulation quantifies the uncertainty arising from Gaussian fields,
or the uncertainty in the {\it absence} of non-Gaussian fluctuations.
The other kind generates non-Gaussian CMB sky maps from primordial 
fluctuations of the form of equation~(\ref{eq:phi})
(see Appendix~\ref{app:simulations} for our method for simulating 
non-Gaussian maps).
This simulation quantifies the uncertainty more accurately and consistently
in the {\it presence} of non-Gaussian fluctuations.

In principle, one should always use the non-Gaussian simulations to 
characterize the uncertainty in $f_{\rm NL}$; however, the uncertainty 
estimated from the Gaussian realizations is good approximation to that from 
the non-Gaussian ones as long as $\left|f_{\rm NL}\right|<500$. 
Our non-Gaussian simulations verify that the distribution of $f_{\rm NL}$
and $b_{\rm src}$ around the mean is the same for Gaussian and non-Gaussian 
realizations (see Figure~\ref{fig:sim_hist} in
Appendix~\ref{app:simulations} for an example of $f_{\rm NL}=100$).
The Gaussian simulations have the advantage of being much faster
than the non-Gaussian ones.
The former takes only a few seconds to simulate one map whereas the 
latter takes 3 hours on a single processor of an SGI Origin 300.
Also, simulating non-Gaussian maps at $nside=512$ requires 17~GB of physical 
memory.
We therefore use Gaussian simulations to estimate the uncertainty in measured
$f_{\rm NL}$ and $b_{\rm src}$.

\subsection{Limits to Primordial Non-Gaussianity}

Figure~\ref{fig:f_NL} shows $f_{\rm NL}$ measured from the Q$+$V$+$W 
coadded map using the cubic statistic
[Eq~(\ref{eq:f_NL})], as a function of the maximum multipole $l_{\rm max}$.
We find the best estimate of $f_{\rm NL}=38\pm 48$ (68\%) for
$l_{\rm max}=265$. The distribution of $f_{\rm NL}$ is close to a
Gaussian, as suggested by Monte Carlo simulations
(see Figure~\ref{fig:sim_hist} in Appendix~\ref{app:simulations}).
The 95\% confidence interval is $-58<f_{\rm NL}<134$.
There is no significant detection of $f_{\rm NL}$ at any angular scale.
The r.m.s. error, estimated from 500 Gaussian simulations, initially 
decreases as $\propto l^{-1}_{\rm max}$, although $f_{\rm NL}$ for 
$l_{\rm max}=265$ has a smaller error than that for $l_{\rm max}=512$ 
because the latter is dominated by the instrumental noise. 
Since all the pixels outside the cut region are uniformly weighted,
the inhomogeneous noise in the map
(pixels on the ecliptic equator are noisier than those on the north and 
south poles) is not accounted for.
This leads to a noisier estimator than a minimum variance estimator. 
The constraint on $f_{\rm NL}$ for $l_{\rm max}=512$ will improve with more 
appropriate pixel-weighting schemes \citep{heavens:1998,santos/etal:2002b}.
The simple inverse noise ($N^{-1}$) weighting makes the constraints 
much worse than the uniform weighting, as it increases errors on 
large angular scales where the CMB signal dominates over the 
instrumental noise.
The uniform weighting is thus closer to optimal.
Note that for the power spectrum, one can simply use the uniform weighting
to measure $C_l$ at small $l$ and the $N^{-1}$ weighting at large $l$.
For the bispectrum, however, this decomposition is not simple, as 
the bispectrum ${\cal B}_{l_1l_2l_3}$ measures the mode coupling from 
$l_1$ to $l_2$ and $l_3$ and {\it vice versa}.
This property makes it difficult to use different weighting schemes 
on different angular scales.
The first column of Table~\ref{tab:values} shows $f_{\rm NL}$
measured in Q, V, and W bands separately. 
There is no a significant band-to-band variation, or a significant detection
in any band.

\begin{figure}
 \epsscale{0.8}
\plotone{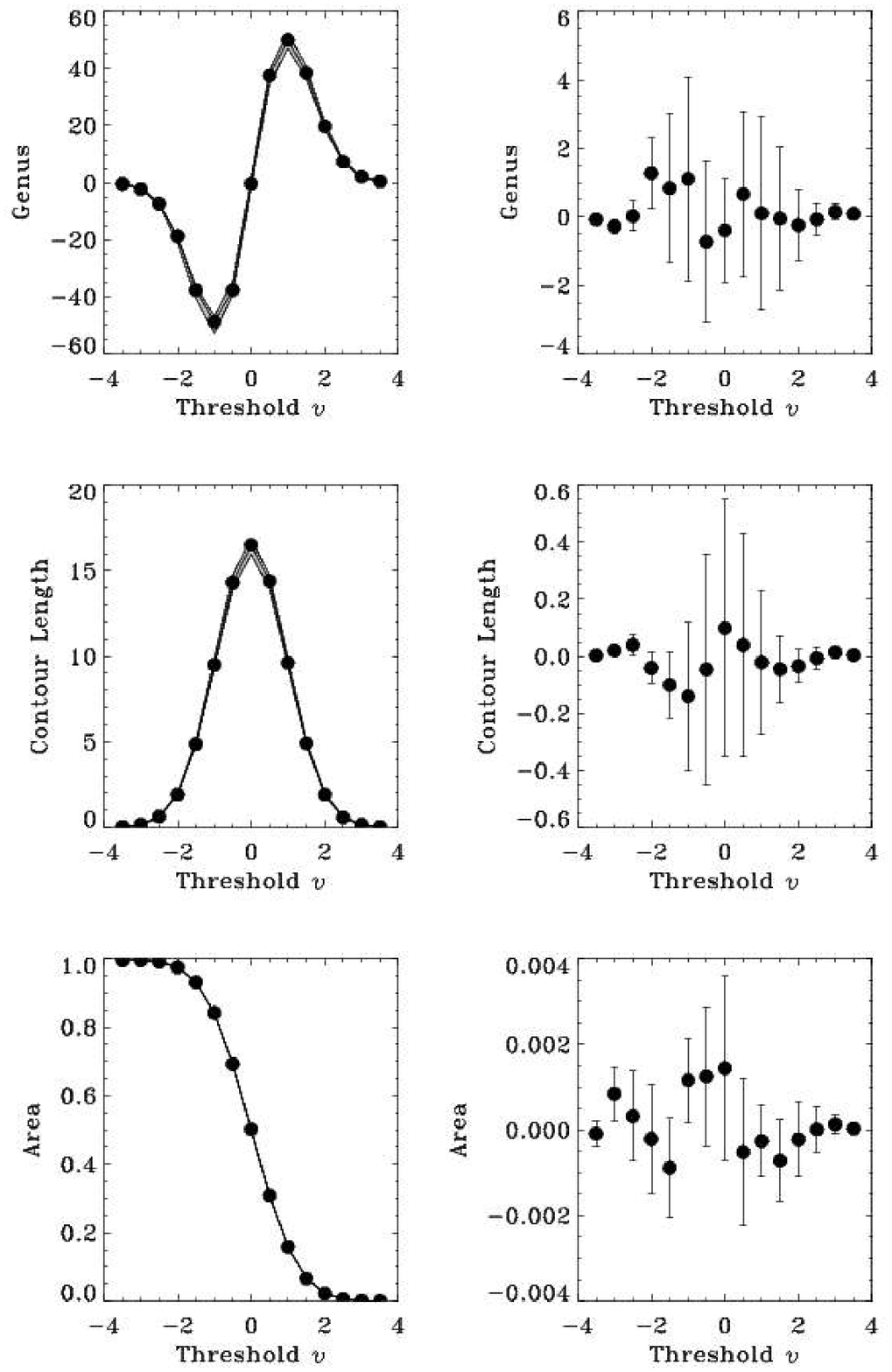}
 \caption{
 The left panels show the Minkowski functionals for {\MAP} data 
 (filled circles) at $nside=128$ (28$\arcmin$ pixels).
 The gray band shows the 68\% confidence interval for the Gaussian 
 Monte Carlo simulations.
 The right panels show the residuals between the mean of the Gaussian
 simulations and the {\MAP} data.
 The {\MAP} data are in excellent agreement with the Gaussian simulations.
 \label{minkowski_res7}
 }
\end{figure}

\begin{figure}
\plotone{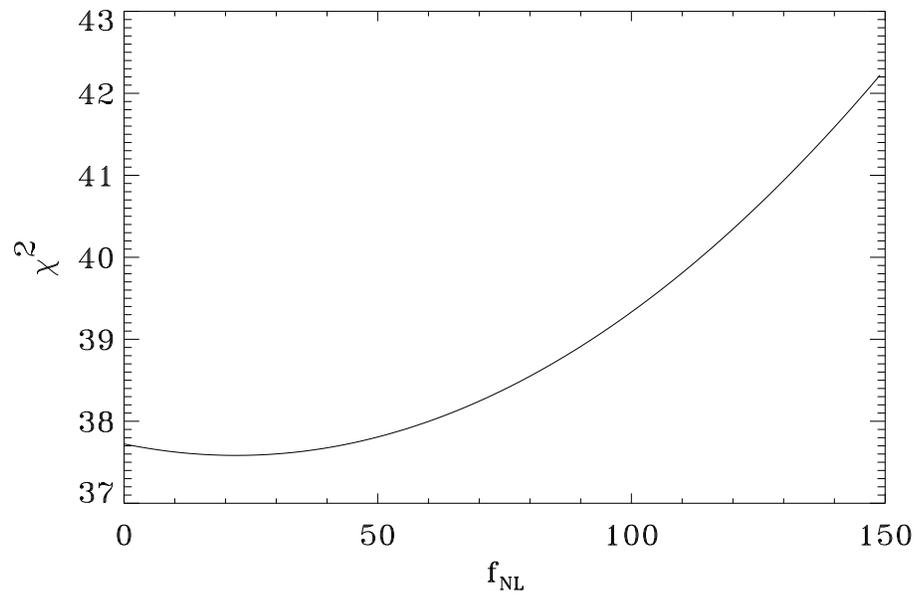}
 \caption{
 Limits to $f_{\rm NL}$ from $\chi^2$ fit of the {\MAP} data to the
 non-Gaussian models [Eq.~(\ref{eq:phi})].
 The fit is a joint analysis of the three Minkowski functionals
 at $28\arcmin$ pixel resolution. There are 44 degrees of freedom.
 \label{minkowski_chi_fig}
 }
\end{figure}

Figure \ref{minkowski_res7} shows the Minkowski functionals
at $nside=128$ (147,594 high-latitude pixels, each $28'$ in diameter).
The gray band shows the 68\% confidence region 
derived from 1000 Gaussian simulations.
Table \ref{minkowsi_chisq_vs_res} shows the $\chi^2$ values 
[Eq.(\ref{chi_def})].
The data are in excellent agreement with the Gaussian simulations
at all resolutions.
The individual Minkowski functionals are highly correlated with each other
(e.g., \citet{shandarin/etal:2002}).
We account for this using a simultaneous analysis
of all three Minkowski functionals, replacing the 15-element vectors
$F^i_{{\rm WMAP},{\nu}}$ and $\langle F^i_{{\rm sim},{\nu}} \rangle$
in equation~(\ref{chi_def}) (the index $i$ denotes each Minkowski
functional) with 45-element vectors
$F_{\nu} = [F^1, F^2, F^3]_{\nu}$ = $[$Area, Contour, Genus$]_{\nu}$
and using the covariance of this larger vector as derived from the simulations.
We compute $\chi^2$ for values $f_{\rm NL}$ = 0 to 1000,
comparing the results from {\MAP} to similar $\chi^2$ values
computed from non-Gaussian realizations.
Figure \ref{minkowski_chi_fig} shows the result.
We find a best-fit value $f_{\rm NL} = 22 \pm 81$ (68\%),
with 95\% confidence upper limit $f_{\rm NL} < 139$,
in agreement with the cubic statistic.

\subsection{Implications of the {\MAP} Limits on $f_{\rm NL}$}

\subsubsection{Inflation}

The limits on $f_{\rm NL}$ are consistent with simple inflation models:
models based on a slowly rolling scalar field typically give
$\left|f_{\rm NL}\right|\sim 10^{-2}-10^{-1}$ 
\citep{salopek/bond:1990,salopek/bond:1991,falk/rangarajan/srendnicki:1993,gangui/etal:1994,acquaviva/etal:2002,maldacena:2002},
three to four orders of magnitude below our limits.
Measuring $f_{\rm NL}$ at this level is difficult because of the cosmic 
variance.
There are alternative models which allow larger amplitudes of 
non-Gaussiantiy in the primordial fluctuations, which we explore below.  

\begin{table}
 \begin{center}
 \caption{
  The non-linear coupling parameter, the reduced point-source angular 
  bispectrum, and the point-source angular power spectrum (positive definite) 
  by frequency band. The errorbars are 68\%. The tabulated values are
  for the {\it Kp0} mask, while the {\it Kp2} mask gives similar 
  results. \label{tab:values}
  }
  \begin{tabular}{cccc}
   \tableline\tableline\noalign{\smallskip}
   & $f_{\rm NL}$ & $b_{\rm src}$ & $c_{\rm src}$ \\
   & & [$10^{-5}~{\rm \mu K^3~sr^2}$] &
   [$10^{-3}~{\rm \mu K^2~sr}$] \\
   \noalign{\smallskip}\tableline\noalign{\smallskip}
   Q 
   & $51\pm 61$ & $9.5\pm 4.4$ & $15\pm 6$ \\
   V
   & $42\pm 63$ & $1.1\pm 1.6$ &  $4.5\pm 4$ \\
   W 
   & $37\pm 75$ & $0.28\pm 1.3$  & --- \\
   Q$+$V$+$W
   & $38 \pm 48$ & $0.94\pm 0.86$ & --- \\
   \noalign{\smallskip}\tableline\tableline\noalign{\smallskip}
  \end{tabular}
 \end{center}
\end{table}

\begin{figure}
\plotone{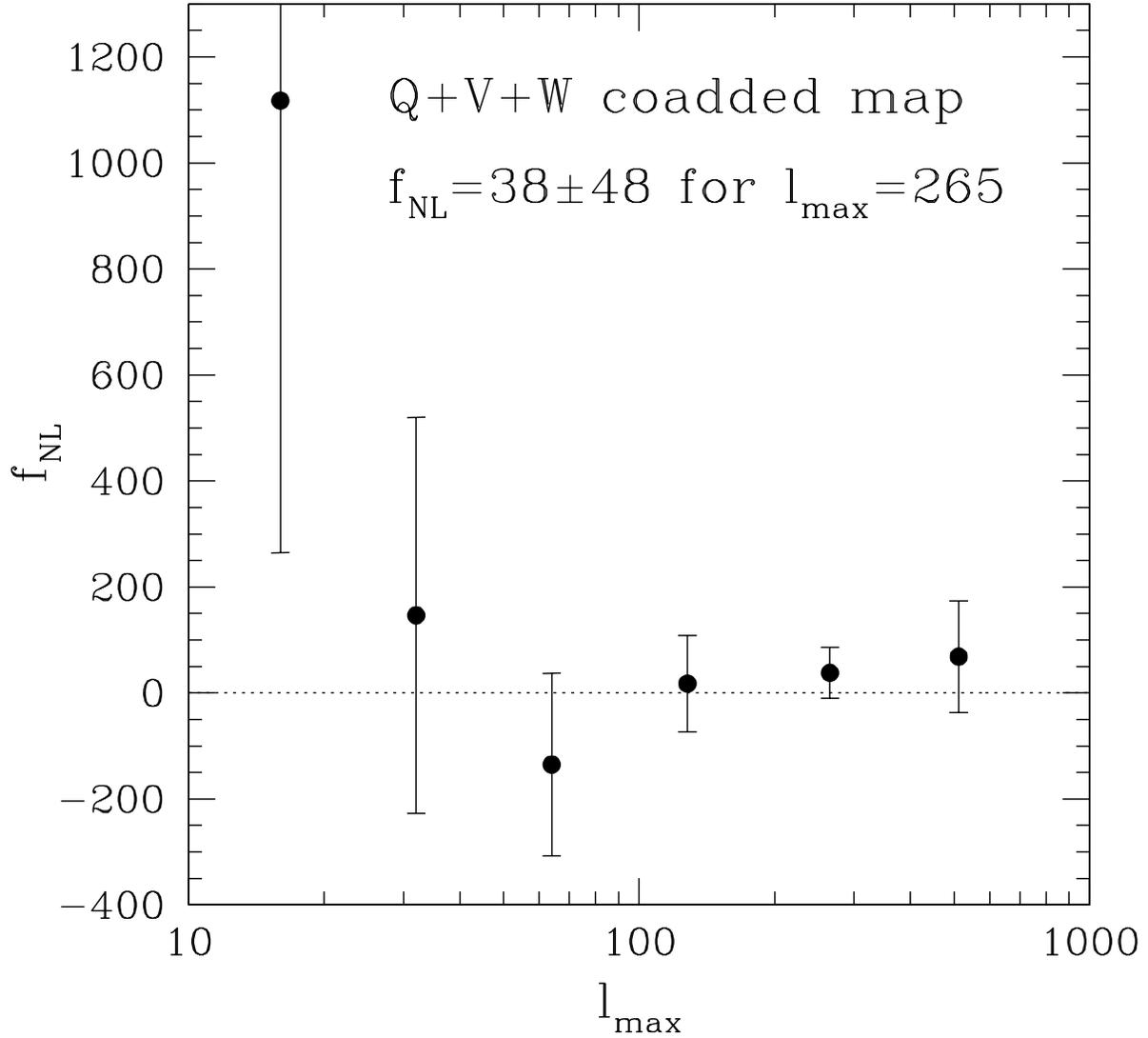}
 \caption{
 The non-linear coupling parameter $f_{\rm NL}$ as a function
 of the maximum multipole $l_{\rm max}$, measured from the Q$+$V$+$W
 coadded map using the cubic (bispectrum) estimator [Eq.~(\ref{eq:f_NL})].
 The best constraint is obtained from $l_{\rm max}=265$. 
 The distribution is cumulative, so that 
 the error bars at each $l_{\rm max}$ are not independent. \label{fig:f_NL}
 }
\end{figure}

A large $f_{\rm NL}$ may be produced when the following condition is met.
Suppose that $\Phi$ is given by $\Phi= \epsilon x$ where $\epsilon$ 
is a transfer function that converts $x$ to $\Phi$ and 
$x=x^{(1)}+x^{(2)}+{\cal O}(x^{(3)})$ 
denotes a fluctuating field expanded into a series of
$x^{(i)}= f_ix^{(i-1)}x^{(1)}$ with $f_1=1$.
Then, $f_{\rm NL}=\epsilon^{-1}f_2$.
Inflation predicts the amplitude of $x^{(i)}$ and the form of $f_i$ which
eventually depends upon the scalar field potential; thus, 
$x^{(i)}$ would be of order $(H/m_{\rm planck})^i$ ($H$ is the Hubble
parameter during inflation) for $H<m_{\rm planck}$, and the leading order 
term is $\epsilon H/m_{\rm planck}\sim 10^{-5}\epsilon$.
In this way $\epsilon$ ``suppresses'' the amplitude of fluctuations, allowing
a larger amplitude for $H/m_{\rm planck}\sim 10^{-5}\epsilon^{-1}$.
What does this mean? 
If $H\sim 10^{-2}m_{\rm planck}$, then $\epsilon\sim 10^{-3}$ and
$f_{\rm NL}\sim 10^3 f_2$.
The amplitude of $f_{\rm NL}$ is thus large enough to detect for $f_2\ga 0.1$.
This suppression factor, $\epsilon$, seems necessary for one to obtain 
a large $f_{\rm NL}$ in the context of the slow-roll inflation. 
The suppression also helps us to avoid a ``fine-tuning problem'' of
inflation models, as it allows $H/m_{\rm planck}$ to be of order slightly
less than unity (which one might think natural) rather than forcing it
to be of order $10^{-5}$.

\begin{table}
 \begin{center}
 \caption{
  $\chi^2$ for Minkowski Functionals\tablenotemark{a}
  \label{minkowsi_chisq_vs_res}
  }
  \begin{tabular}{c c c c c}
  \tableline 
  $nside$ & Pixel Diam  & Minkowski     & {\MAP}                
  & $f(>${\MAP}$)$\tablenotemark{b} \\
         & (deg)        & Functional    & $\chi^2$      & \\
  \tableline 
   256   &   0.2   &   Genus      &   15.9     &    0.57 \\
   128   &   0.5   &   Genus      &   10.7     &    0.79 \\
    64   &   0.9   &   Genus      &   15.7     &    0.44 \\
    32   &   1.8   &   Genus      &   18.7     &    0.26 \\
    16   &   3.7   &   Genus      &   16.8     &    0.22 \\
         &         &              &            &         \\
   256   &   0.2   &   Contour    &    9.9     &    0.93 \\
   128   &   0.5   &   Contour    &    9.9     &    0.83 \\
    64   &   0.9   &   Contour    &   14.6     &    0.54 \\
    32   &   1.8   &   Contour    &   12.8     &    0.58 \\
    16   &   3.7   &   Contour    &   11.9     &    0.67 \\
         &         &              &            &         \\
   256   &   0.2   &   Area       &   17.4     &    0.50 \\
   128   &   0.5   &   Area       &   10.9     &    0.74 \\
    64   &   0.9   &   Area       &   11.9     &    0.66 \\
    32   &   1.8   &   Area       &   21.9     &    0.12 \\
    16   &   3.7   &   Area       &   15.7     &    0.33 \\
  \tableline 
  \end{tabular}
  \tablenotetext{a}{$\chi^2$ computed using Gaussian simulations.
  There are 15 degrees of freedom.}
  \tablenotetext{b}{Fraction of simulations with $\chi^2$ greater than
  the value from the {\MAP} data.}
 \end{center}
\end{table}

Curvatons proposed by \cite{lyth/wands:2002} provide an example of
a supression mechanism.
A curvaton is a scalar field, $\sigma$, having mass, $m_\sigma$, that develops 
fluctuations, $\delta\sigma$, during inflation with its 
energy density, $\rho_\sigma\simeq V(\sigma)$, tiny compared to that 
of the inflaton field that drives inflation.
After inflation ends, radiation is produced as the inflaton decays, generating
entropy perturbations between $\sigma$ and radiation, 
$S_{\sigma\gamma}=\delta\rho_\sigma/\rho_\sigma 
- \frac34\delta\rho_\gamma/\rho_\gamma$. 
When $H$ decreases to become comparable to $m_\sigma$, 
oscillations of $\sigma$ at the bottom of $V(\sigma)$ give
$\rho_\sigma\simeq m_\sigma^2\sigma^2$.
In the limit of ``cold inflation'' for which $\delta\rho_\gamma/\rho_\gamma$
is nearly zero, one finds $S_{\sigma\gamma}\simeq 
\delta\rho_\sigma/\rho_\sigma\simeq
2\delta\sigma/\sigma + (\delta\sigma/\sigma)^2$.
As long as $\sigma$ survives after the production of $S_{\sigma\gamma}$,
the curvature perturbation $\Phi$ is generated as $\Phi=\frac12\epsilon 
S_{\sigma\gamma}\simeq \epsilon[x^{(1)}+\frac12(x^{(1)})^2]$ where 
$x^{(1)}=\delta\sigma/\sigma$ (i.e., $f_2=\frac12$).
The generation of $\Phi$ continues until $\sigma$ decays, and
$\Phi$ is essentially determined by a ratio of $\rho_\sigma$ to the total
energy density, $\Omega_\sigma$, at the time of the decay.
\cite{lyth/ungarelli/wands:2002} numerically evolved perturbations to
find $\epsilon\simeq \frac25\Omega_\sigma$ at the time of the decay.
The smaller the curvaton energy density is, the less efficient the 
$S_{\sigma\gamma}$ to $\Phi$ conversion becomes
(or the more efficient the supression becomes).
The small $\Omega_\sigma$ thus leads to the large $f_{\rm NL}$, as 
$f_{\rm NL}=\epsilon^{-1}f_2\simeq \frac54\Omega_\sigma^{-1}$
(i.e., $f_{\rm NL}$ is always positive in this model).
Assuming the curvaton exists and is entirely responsible for the observed 
CMB anisotropy, our limits on $f_{\rm NL}$ imply
$\Omega_{\sigma}>9\times 10^{-3}$ at the time of the curvaton decay.
(However, the lower limit to $\Omega_{\sigma}$ does not mean that we 
need the curvatons. This constraint makes sense only when the curvaton
exists and is entierly producing the observed fluctuations.)

Features in an inflaton potential can generate significant non-Gaussian
fluctuations \citep{kofman/etal:1991,wang/kamionkowski:2000},
and it is expected that measurements of non-Gaussianity can place constrains
on a class of the feature models.
In Appendix~\ref{app:step}, we calculate the angular bispectrum
from a sudden step in a potential of the form in equation~(\ref{eq:feature}).
This step is motivated by a class of supergravity models yielding
the steps as a consequence of successive spontaneous symmetry-breaking 
phase transitions of many fields coupled to the inflaton
\citep{adams/ross/sarkar:1997b,adams/cresswell/easther:2001}.
One step generates two distinct regions in $l$ space where 
$\left|f_{\rm NL}\right|$ is very large: a positive $f_{\rm NL}$ is 
predicted at $l<l_{\rm f}$, while a negative $f_{\rm NL}$ at $l>l_{\rm f}$,
where $l_{\rm f}$ is the projected location of the step.
Our calculations suggest that the two regions are separated in $l$ by 
less than a factor of 2, and one cannot resolve them without knowing 
$l_{\rm f}$.
The average of many $l$ modes further smeares out the signals.
The averaged $f_{\rm NL}$ thus nearly cancels out to give only small 
signals, being hidden in our constraints in Figure~\ref{fig:f_NL}.
\cite{peiris/etal:2003} argue that some sharp features in the
{\MAP} angular power spectrum producing large $\chi^2$ values 
may arise from features in the inflaton potential. 
If this is true, then one may be able to see non-Gaussian signals
associated with the features by measuring the bispectrum at the 
scales of the sharp features of the power spectrum.

\subsubsection{Massive cluster abundance at high redshift}

\begin{figure}
\plotone{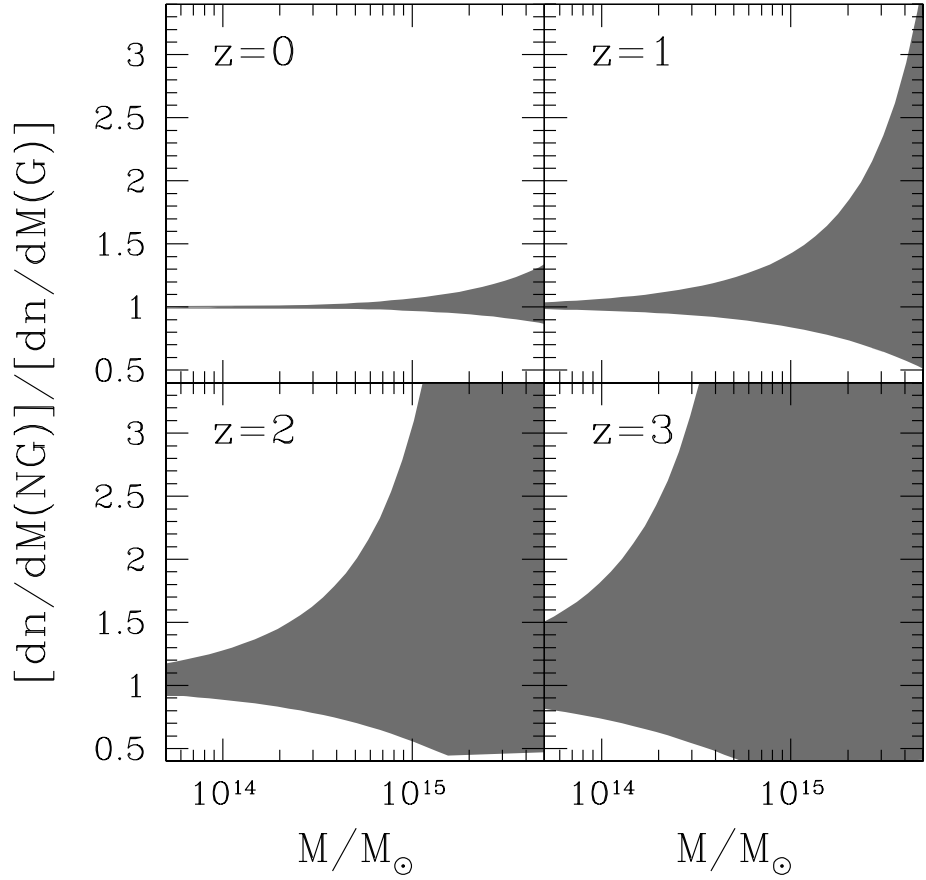}
 \caption{
 The limits to the effect of the primordial non-Gaussianity on
 the dark-matter halo mass function $dn/dM$ as a function of $z$.
 The shaded area represents the 95\% constraint on the ratio of 
 the non-Gaussian $dn/dM$ to the Gaussian one.
 \label{fig:dndm}
 }
\end{figure}

\begin{figure}
\plotone{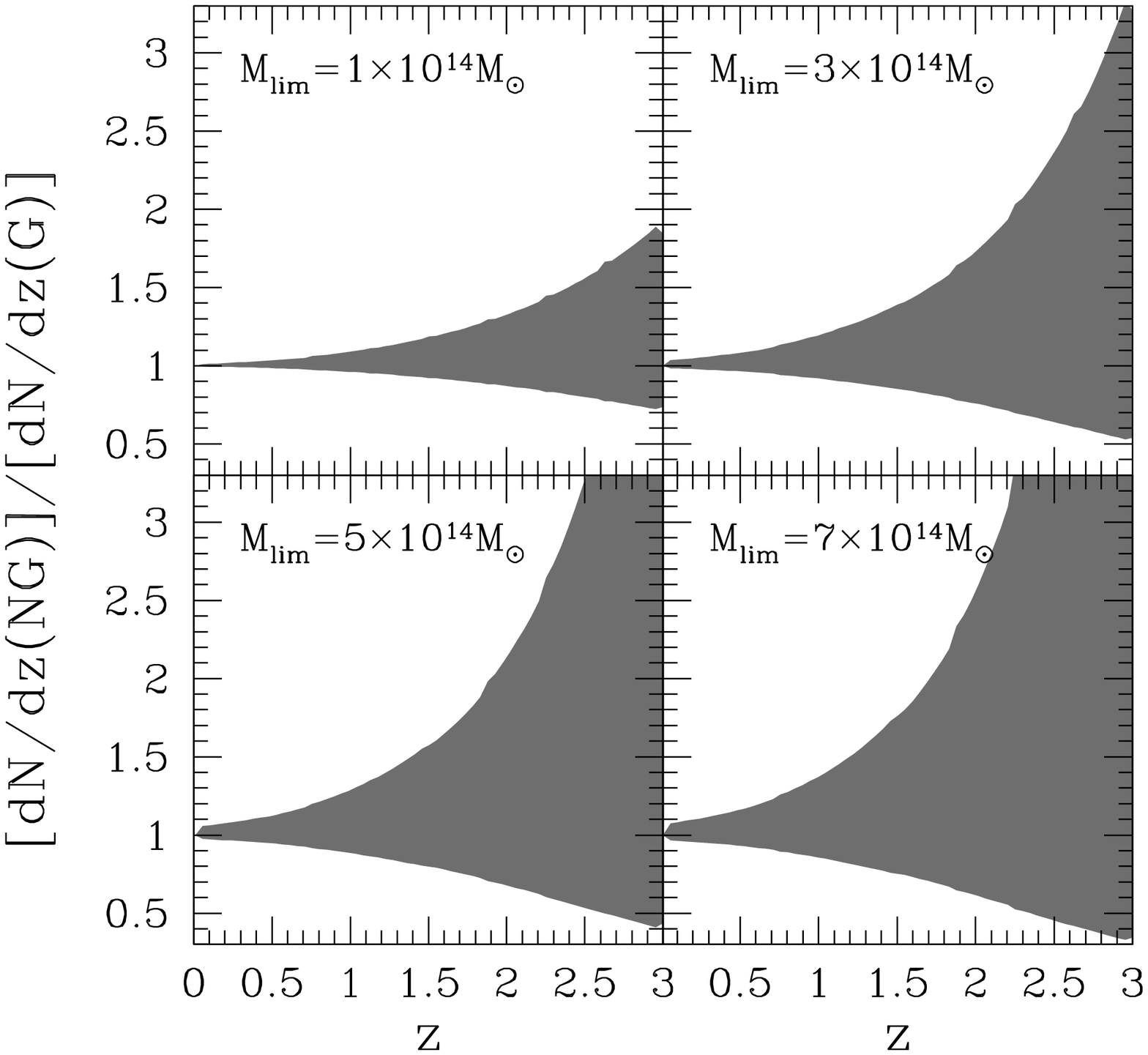}
 \caption{
 The same as figure~\ref{fig:dndm} but for the dark-matter halo
 number counts $dN/dz$ as a function of the limiting mass $M_{\rm lim}$
 of a survey. \label{fig:dNdz}
 }
\end{figure}

Massive halos, like clusters of galaxies at high redshift, are such rare 
objects in the universe that their abundance is sensitive to the presence
of non-Gaussianity in the distribution function of primordial density 
fluctuations.
Several authors have pointed out the power of the halo abundance as a tool for
finding primordial non-Gaussianity
\citep{lucchin/matarrese:1988,robinson/baker:2000,matarrese/verde/jimenez:2000,benson/reichardt/kamionskowski:2002}; however, the power of this method is extremely sensitive to 
the accuracy of the mass determinations of halos.
It is necessary to go to redshifts of $z\ga 1$ to obtain tight
constraints on primordial non-Gaussianity, as
constraints from low and intermediate redshifts appear to be
weak \citep{koyama/soda/taruya:1999,robinson/gawiser/silk:2000} 
(see also Figure~\ref{fig:dndm} and \ref{fig:dNdz}).
Due to the difficulty of measuring the mass of a high-redshift cluster 
the current constraints are not yet conclusive \citep{willick:2000}.
The limited number of clusters observed at high redshift also limits
the current sensitivity.
In this section, we translate our constraints on $f_{\rm NL}$ from 
the {\MAP} 1-year CMB data into the effects on the massive halos in 
the high-redshift universe, showing the extent to which future cluster 
surveys would see signatures of non-Gaussian fluctuations.

We adopt the method of \cite{matarrese/verde/jimenez:2000} to
calculate the dark-matter halo mass function $dn/dM$ 
for a given $f_{\rm NL}$, using the $\Lambda$CDM with
the running spectral index model best-fit to the {\MAP} data and 
the large-scale structure data.
This set of parameters is best suited for the calculations of 
the cluster abundance.
The parameters are in the rightmost column of Table~8 of 
\cite{spergel/etal:2003}.
We calculate
\begin{equation}
 \label{eq:ngmf}
 \frac{dn}{dM}= 2\frac{\rho_{\rm m0}}M\left|\frac{dP}{dM}\right|,
\end{equation}
where $\rho_{\rm m0}=2.775\times 10^{11} 
(\Omega_{\rm m}h^2)~M_\odot~{\rm Mpc^{-3}}
= 3.7\times 10^{10}~M_\odot~{\rm Mpc^{-3}}$
is the present-day mean mass density of the 
universe, $P(M,z)$ is the probability for halos of mass $M$ to collapse
at redshift $z$, and $dP/dM$ is given by
\begin{equation}
 \label{eq:dpdm}
 \frac{dP}{dM} \equiv \int_0^\infty \frac{\lambda d\lambda}{2\pi}
 \left(
 \frac{d\sigma^2}{dM} \sin\theta_\lambda
 -\frac{\lambda}3 \frac{d\mu_3}{dM}\cos\theta_\lambda
 \right)
 e^{-\lambda^2\sigma^2/2},
\end{equation}
where the angle $\theta_\lambda$ is given by 
$\theta_\lambda \equiv \lambda\delta_{\rm c} + \lambda^3\mu_3/6$, and
$\delta_c(z)$ is the threshold overdensity of spherical collapse
\citep{lacey/cole:1993,nakamura/suto:1997}.
The variance of the mass fluctuations as a function of $z$ is 
given by 
$\sigma^2(M,z)=D^2(z)\sigma^2(M,0)$, where
$D(z)$ is the growth factor of linear density fluctuations,
$\sigma^2(M,0)=\int_0^\infty dkk^{-1} F^2_M(k) \Delta^2(k)$,
$\Delta^2(k)\equiv (2\pi^2)^{-1} k^3 P(k)$ is the 
dimensionless power spectrum of the Bardeen curvature perturbations,
$F_M(k)\equiv g(k)T(k)W(kR_M)$ a filter function,
$g(k)\equiv \frac23(k/H_0)^2\Omega_{\rm m0}^{-1}$
a conversion factor from $\Phi$ to density fluctuations,
$T(k)$ the transfer function of linear density perturbations, 
$W(x)\equiv 3j_1(x)/x$ the spherical top-hat window smoothing density fields, 
and 
$R_M\equiv [3M/(4\pi\rho_{\rm m0})]^{1/3}$ the spherical top-hat radius 
enclosing a mass $M$.
The skewness $\mu_3(M,z)=D^3(z)\mu_3(M,0)$, where
\begin{equation}
 \label{eq:mu3}
 \mu_3(M,0)= 6 f_{\rm NL}
 \int_0^\infty \frac{dk_1}{k_1} F_M(k_1) \Delta^2(k_1)
 \int_0^\infty \frac{dk_2}{k_2} F_M(k_2) \Delta^2(k_2)
 \int_0^1 d\mu F_M(\sqrt{k_1^2+k_2^2+2k_1k_2\mu}),
\end{equation}
arises from the primordial non-Gaussianity.
We use a Monte Carlo integration routine called {\sf vegas} 
\citep{press/etal:NRIC:2e} to 
evaluate the triple integral in equation~(\ref{eq:mu3}).
It follows from equation~(\ref{eq:mu3}) that a positive $f_{\rm NL}$
gives a positive $\mu_3$, positively skewed density fluctuations.
Also this $dn/dM$ reduces to the Press--Schechter form
\citep{press/schechter:1974} in the limit of $f_{\rm NL}\rightarrow 0$.
Although the Press--Schechter form predicts significantly fewer 
massive halos than $N$-body simulations \citep{jenkins/etal:2001},
we assume that a predicted ratio of the non-Gaussian $dn/dM$ to the Gaussian
$dn/dM$ is still reasonably accurate, as the primordial non-Gaussianity 
does not affect the dynamics of halo formations which causes
the difference between the Press--Schechter form of $dn/dM$ 
and the $N$-body simulations.

Figure~\ref{fig:dndm} shows the {\MAP} constraints on the ratio of 
non-Gaussian $dn/dM$ to the Gaussian one, as a function of $M$ and $z$.
We find that the {\MAP} constraint on $f_{\rm NL}$
strongly limits the amplitude of changes in $dn/dM$ due to 
the non-Gaussianity.
At $z=0$, $dn/dM$ is changed by no more than 20\% even for 
$4\times 10^{15}~M_\odot$ clusters.
The number of clusters that would be newly found
at $z=1$ for $M<10^{15}~M_\odot$ should be within ${}^{+40}_{-10}\%$
of the value predicted from the Gaussian theory.
At $z=3$, however, much larger effects are still allowed: 
$dn/dM$ can be increased by up to a factor of 2.5 for 
$2\times 10^{14}~M_\odot$.

Predictions for actual cluster surveys are made clearer by
computing the source number counts as a function of $z$, 
\begin{equation}
 \label{eq:dNdz}
  \frac{dN}{dz}\equiv \frac{dV}{dz}\int_{M_{\rm lim}}^\infty dM
  \frac{dn}{dM},
\end{equation}
where $V(z)$ is the comoving volume per steradian, and $M_{\rm lim}$
is the limiting mass that a survey can reach.
In practice $M_{\rm lim}$ would depend on $z$ due to, for example,
the redshift dimming of X-ray surface brightness; however,
a constant $M_{\rm lim}$ turns out to be a good 
approximation for surveys of the Sunyaev--Zel'dovich (SZ) effect 
\citep{carlstrom/holder/reese:2002}.
Figure~\ref{fig:dNdz} shows the ratio for $dN/dz$ as a function of
$z$ and $M_{\rm lim}$.
A source-detection sensitivity of 
$S_{\rm lim}=0.5~{\rm Jy}$ roughly corresponds to 
$M_{\rm lim}= 1.4\times 10^{14}~M_\odot$ 
\citep{carlstrom/holder/reese:2002}, for which 
$dN/dz$ should follow the prediction of the Gaussian
theory out to $z\simeq 1$ to within 10\%, but
$dN/dz$ at $z=3$ can be increased by up to a factor of 2.
As $M_{\rm lim}$ increases, the impact on $dN/dz$ rapidly increases.

The SZ angular power spectrum $C_l^{\rm SZ}$ is so sensitive to 
$\sigma_8$ that we can use $C_l^{\rm SZ}$ to measure $\sigma_8$ 
\citep{komatsu/kitayama:1999}.
The sensitivity arises largely from massive ($M>10^{14}~M_\odot$)
clusters at $z\sim 1$.
From this fact one might argue that $C_l^{\rm SZ}$ is also 
sensitive to the primordial non-Gaussianity.
We use a method of \cite{komatsu/seljak:2002} with $dn/dM$
replaced by equation~(\ref{eq:ngmf}) to compute $C_l^{\rm SZ}$
for the {\MAP} limits on $f_{\rm NL}$.
We find that $C_l^{\rm SZ}$ should follow the prediction from the
Gaussian theory to within 10\% for $100<l<10000$.
This is consistent with $C_l^{\rm SZ}$ being primarily sensitive to
halos at $z\sim 1$, where the effect on $dN/dz$ is not too strong 
(see Figure~\ref{fig:dNdz}).
Since $C_l^{\rm SZ}\propto \sigma_8^7(\Omega_{\rm b}h)^2$ 
\citep{komatsu/seljak:2002}, $\sigma_8$ can be determined 
from $C_l^{\rm SZ}$ to within 2\% accuracy at a fixed $\Omega_{\rm b}h$
using the Gaussian theory.
The current theoretical uncertainty in the predictions of $C_l^{\rm SZ}$
is a factor of 2 in $C_l^{\rm SZ}$ (10\% in $\sigma_8$), still much larger 
than the effect of the non-Gaussianity.

\section{LIMITS TO RESIDUAL POINT SOURCES}\label{sec:pointsources}

\subsection{Point-source Angular Power Spectrum and Bispectrum}

Radio point sources distributed across the sky generate 
non-Gaussian signals, giving a positive bispectrum, $b_{\rm src}$
\citep{komatsu/spergel:2001}.
In addition, the point sources contribute significantly to the angular 
spectrum on small angular scales \citep{tegmark/efstathiou:1996}, 
contaminating the cosmological angular power spectrum.
It is thus important to understand how much of the measured angular power spectrum
is due to sources.
We constrain the source contribution to the angular power spectrum, 
$c_{\rm src}$, by measuring $b_{\rm src}$. 
\cite{komatsu/spergel:2001} have shown that {\MAP} can detect 
$b_{\rm src}$ even after subtracting all (bright) sources detected in 
the sky maps. 
Fortunately, there is no degeneracy between $f_{\rm NL}$ 
and $b_{\rm src}$, as shown later in Appendix~\ref{app:simulations}.

In this section we measure the amplitude of non-Gaussianity 
from ``residual'' point sources which are fainter than a certain 
flux threshold, $S_{\rm c}$, and left unmasked in the sky maps.
The bispectrum $b_{\rm src}$ is related to the number of sources brighter than 
$S_{\rm c}$ per solid angle $N(>S_{\rm c})$:
\begin{equation}
 \label{eq:bsrc}
  b_{\rm src}(S_{\rm c})= \int_0^{S_{\rm c}}dS \frac{dN}{dS} [g(\nu)S]^3
  = -N(>S_{\rm c}) [g(\nu)S_{\rm c}]^3 
    + 3\int_0^{S_{\rm c}} \frac{dS}S N(>S) [g(\nu)S]^3,
\end{equation}
where $g(\nu)$ is a conversion factor from ${\rm Jy~sr^{-1}}$ to $\mu$K  
which depends on observing frequency $\nu$ as
$g(\nu) = (24.76~{\rm Jy~\mu K^{-1}~sr^{-1}})^{-1}[(\sinh x/2)/x^2]^2$, 
$x\equiv h\nu/k_{\rm B}T_0\simeq \nu/(56.78~{\rm GHz})$
for $T_0=2.725~{\rm K}$ \citep{mather/etal:1999}, 
and $dN/dS$ is the differential source count per solid angle.
The residual point sources also contribute to the point-source 
power spectrum $c_{\rm src}$ as
\begin{equation}
 \label{eq:csrc}
  c_{\rm src}(S_{\rm c})= \int_0^{S_{\rm c}}dS \frac{dN}{dS} [g(\nu)S]^2
  = -N(>S_{\rm c}) [g(\nu)S_{\rm c}]^2
    + 2\int_0^{S_{\rm c}} \frac{dS}S N(>S) [g(\nu)S]^2.
\end{equation}
By combining equation~(\ref{eq:bsrc}) and (\ref{eq:csrc}) 
we find a relation between $b_{\rm src}$ and $c_{\rm src}$,
\begin{equation}
 \label{eq:bsrc-csrc}
  c_{\rm src}(S_{\rm c})= b_{\rm src}(S_{\rm c})[g(\nu)S_{\rm c}]^{-1}
  +\int_0^{S_{\rm c}} \frac{dS}S b_{\rm src}(S) [g(\nu)S]^{-1}.
\end{equation}
We can use this equation combined with the measured $b_{\rm src}$ as a 
function of $S_{\rm c}$ to directly determine $c_{\rm src}$ as a function of 
$S_{\rm c}$, without relying on any extrapolations.
When the source counts obey a power-law like $dN/dS\propto S^\beta$,
one finds $b_{\rm src}(S)\propto S^{4+\beta}$; thus, brighter sources
contribute more to the integral in equation~(\ref{eq:bsrc-csrc})
than fainter ones as long as $\beta>-3$, which is the case for fluxes 
of interest. 
\cite{bennett/etal:2003c} have found $\beta=-2.6\pm 0.2$ for $S=2-10$~Jy 
in Q band.
Below 1~Jy, $\beta$ becomes even flatter \citep{toffolatti/etal:1998},
implying that one does not have to go down to the very faint end to obtain 
reasonable estimates of the integral.
In practice, we use equation~(\ref{eq:bsrc}) with $N(>S)$ of 
the \cite{toffolatti/etal:1998} model (hereafter T98) at 44~GHz to 
compute $b_{\rm src}(S<0.5~{\rm Jy})$, 
inserting it into the integral to avoid missing faint sources and 
underestimating the integral. 

\subsection{Measurement of the Point-source Angular Bispectrum}

The reduced point-source angular bispectrum, $b_{\rm src}$, is measured 
by a cubic statistic for point sources \citep{komatsu/spergel/wandelt:2003},
\begin{equation}
 \label{eq:skewness*}
  {\cal S}_{ps} =
  m_3^{-1} \int \frac{d^2\hat{\mathbf n}}{4\pi} D^3(\hat{\mathbf n}),
\end{equation}
where the filtered map $D(\hat{\mathbf n})$ is given by
\begin{equation}
 \label{eq:filter3}
 D(\hat{\mathbf n}) \equiv \sum_{l=2}^{l_{\rm max}} \sum_{m=-l}^l
 \frac{b_l}{\tilde{C}_l} a_{lm} Y_{lm}(\hat{\mathbf n}).
\end{equation}
This statistic is even quicker ($\sim 100$ times) to compute 
than $S_{\rm prim}$ (eq.[\ref{eq:skewness}]), as it involves only 
one integral over $\hat{\mathbf n}$ and only one filtered map.
This statistic also retains the same sensitivity to the point-source 
non-Gaussianity as the full bispectrum analysis.
The cubic statistic $S_{\rm ps}$ gives $b_{\rm src}$ as
\begin{equation}
 \label{eq:b_ps*}
  b_{\rm src}
  \simeq 
  \left[
  \frac{3}{2\pi} \sum^{l_{\rm max}}_{l_1\le l_2\le l_3}
  \frac{({\cal B}_{l_1l_2l_3}^{\rm ps})^2}
  {{\cal C}_{l_1}{\cal C}_{l_2}{\cal C}_{l_3}}\right]^{-1}
  {\cal S}_{\rm ps},
\end{equation}
where ${\cal B}_{l_1l_2l_3}^{\rm ps}$ is the point-source bispectrum
for $b_{\rm src}=1$ \citep{komatsu/spergel:2001} 
multiplied by $b_{l_1}b_{l_2}b_{l_3}$. 
While the uniform pixel-weighting outside the Galactic cut was used
for $f_{\rm NL}$, we use here 
$M(\hat{\mathbf n})= \left[\sigma_{\rm CMB}^2+N(\hat{\mathbf n})\right]^{-1}$
where $\sigma_{\rm CMB}^2=(4\pi)^{-1}\sum_l (2l+1)C_lb_l^2$ is the variance
of CMB anisotropy and $N(\hat{\mathbf n})$ is the variance of noise per pixel
which varies across the sky.
This weighting scheme is nearly optimal for measuring $b_{\rm src}$ as
the signal comes from smaller angular scales where noise dominates.
The factor of $\sigma_{\rm CMB}^2$ approximately takes into account 
the non-zero contribution to the variance from CMB anisotropy.
This weight reduces uncertainties of $b_{\rm src}$ by 17\%, 23\%, and 31\% in
Q, V, and W bands, respectively, compared to the uniform weighting.
We use the highest resolution level, $nside=512$, and integrate
equation~(\ref{eq:b_ps*}) up to $l_{\rm max}=1024$.
In Appendix~\ref{app:ptsrc}, it is shown that this estimator is optimal 
and unbiased as long as very bright sources, which have 
contributions to $\tilde{C}_l$ too large to ignore, are masked. 
We cannot include $c_{\rm src}$ in the filter, 
as it is what we are trying to measure using $b_{\rm src}$.

\begin{figure}
\plotone{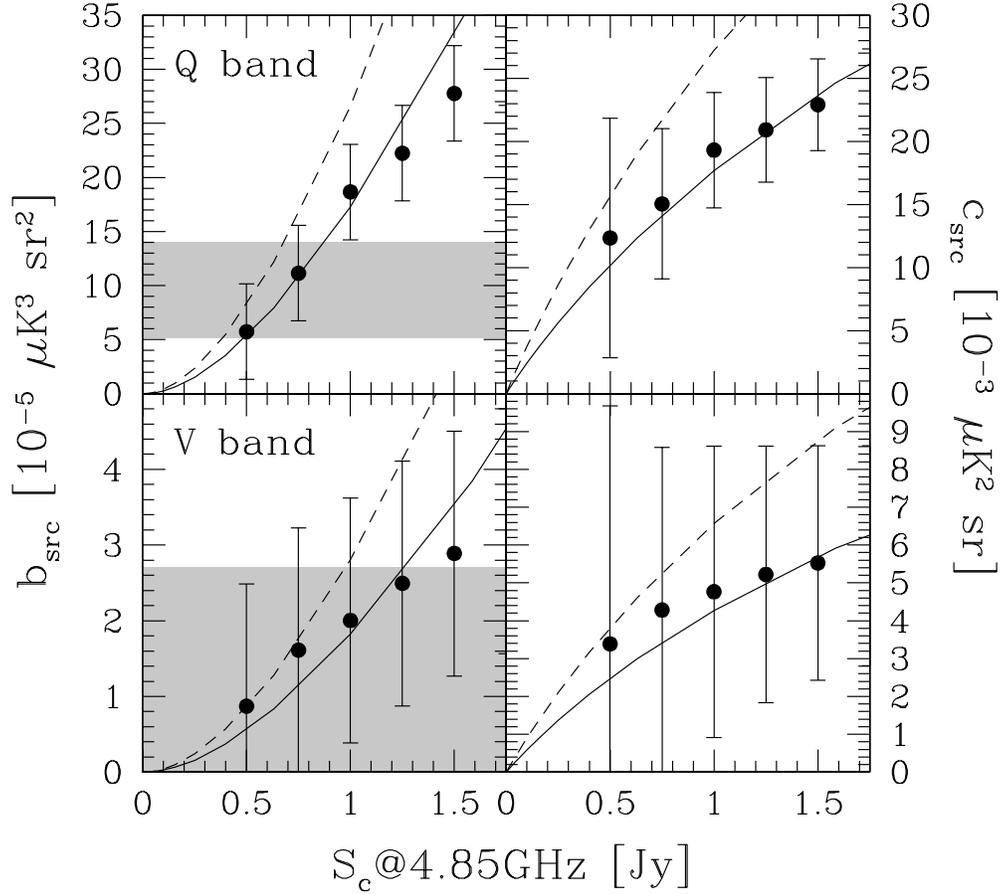}
 \caption{
 The point-source angular bispectrum $b_{\rm src}$ and 
 power spectrum $c_{\rm src}$.
 The left panels show $b_{\rm src}$ in Q band (top panel) and 
 V band (bottom panel).
 The shaded areas show measurements from the {\MAP} sky maps with
 the standard source cut, while the filled circles show those with 
 flux thresholds $S_{\rm c}$ defined at 4.85~GHz. 
 The dashed lines show predictions from equation~(\ref{eq:bsrc}) with
 $N(>S)$ modeled by \cite{toffolatti/etal:1998}, while the solid lines
 are those multiplied by 0.65 to match the {\MAP} measurements.
 The right panels show $c_{\rm src}$.
 The filled circles are computed from the measured $b_{\rm src}$
 substituted into equation~(\ref{eq:bsrc-csrc}).
 The lines are from equation~(\ref{eq:csrc}). \label{fig:ptsrc}
 The error bars are not independent, because the distribution is
 cumulative.
 }
\end{figure}

The filled circles in the left panels of Figure~\ref{fig:ptsrc} represent
$b_{\rm src}$ measured in Q (top panel) and V (bottom panel) band.
We have used source masks for various flux cuts, $S_{\rm c}$, defined at
4.85~GHz to make these measurements. 
(The masks are made from the GB6$+$PMN~5~GHz source catalogue.)
We find that $b_{\rm src}$ increases as $S_{\rm c}$: the brighter sources
unmasked, the more non-Gaussianity is detected.
On the other hand one can make predictions for $b_{\rm src}$ using
equation~(\ref{eq:bsrc}) for a given $N(>S)$.
Comparing the measured values of $b_{\rm src}$ with the predicted values from
$N(>S)$ of T98 (dashed lines) at 44~GHz, one finds that the 
measured values are smaller than the predicted values by a factor of 0.65.
The solid lines show the predictions multiplied by 0.65.
Both errors in the T98 predictions and a non-flat energy 
spectrum of sources easily cause this factor.
(If sources have a non-flat spectrum like 
$S\propto \nu^\alpha$ where $\alpha\neq 0$, then $S_{\rm c}$ at Q or 
V band is different from that at 4.85~GHz.)
\cite{bennett/etal:2003c} find that the majority of the radio sources 
detected in Q band have a flat spectrum, $\alpha=0.0\pm 0.2$.
Our value for the correction factor matches well the one obtained 
from the {\MAP} source counts for $2-10~{\rm Jy}$ in Q band 
\citep{bennett/etal:2003c}.

Equation~(\ref{eq:csrc}) combined with the measured 
$b_{\rm src}$ is used to estimate the point-source angular power spectrum 
$c_{\rm src}$.
The right panels of Figure~\ref{fig:ptsrc} show the estimated
$c_{\rm src}$ as filled circles.
These estimates agree well with predictions from equation~(\ref{eq:csrc})
with $N(>S)$ of T98 multiplied by a factor of 0.65 (solid lines).
For $S_{\rm c}=1~{\rm Jy}$ at Q band, 
$\hat{c}_{\rm src}=(19\pm 5)\times 10^{-3}~\mu{\rm K^2~sr}$, and
matches well the value estimated from the {\MAP} source counts at
the same flux threshold \citep{bennett/etal:2003c}, which corresponds 
to the solid lines in the figure.
At V band, $\hat{c}_{\rm src}=(5\pm 4)\times 10^{-3}~\mu{\rm K^2~sr}$.
Here, the hat denotes that these values do {\it not} represent 
$c_{\rm src}$ for the standard source mask used by \cite{hinshaw/etal:2003}
for estimating the cosmological angular power spectrum. 
Since the standard source mask is made of several source catalogues 
with different selection thresholds, it is difficult to clearly identify
a mask flux cut.
We give the standard mask an ``effective'' flux cut threshold at 4.85~GHz by 
comparing $b_{\rm src}$ measured from the standard source mask 
(shaded areas in Figure~\ref{fig:ptsrc}; see the second column
of Table~\ref{tab:values} for actual values) with those from the
GB6$+$PMN masks defined at 4.85~GHz.
The measurements agree when $S_{\rm c}\simeq 0.75~{\rm Jy}$ in Q band.
Using this effective threshold, one expects $c_{\rm src}$ for the standard
source mask as
$c_{\rm src}=(15\pm 6)\times 10^{-3}~\mu{\rm K^2~sr}$ in Q band.
This value agrees with the excess power seen on small angular 
scales, $(15.5\pm 1.7)\times 10^{-3}~\mu{\rm K}^2~{\rm sr}$ 
\citep{hinshaw/etal:2003}, as well as the value extrapolated
from the {\MAP} source counts in Q band,
$(15.0\pm 1.4)\times 10^{-3}~\mu{\rm K}^2~{\rm sr}$ \citep{bennett/etal:2003c}.
In V band, $c_{\rm src}=(4.5\pm 4)\times 10^{-3}~\mu{\rm K^2~sr}$.

The source number counts, angular power spectrum, and bispectrum
measure the first-, second-, and third-order moments of $dN/dS$,
respectively.
The good agreement among these three different estimates of 
$c_{\rm src}$ indicates the validity of the estimate of the effects of 
the residual point sources in Q band.
There is no visible contribution to the angular power spectrum 
from the sources in V and W bands. 
We conclude that our understanding of the amplitude of the residual 
point sources is satisfactory for the analysis of the angular power spectrum
not to be contaminated by the sources.

\section{CONCLUSIONS}\label{sec:conclusion}

We use cubic (bispectrum) statistics and the Minkowski functionals to 
measure non-Gaussian fluctuations in the {\MAP} 1-year sky maps.
The cubic statistic [Eq.(\ref{eq:skewness})] and the Minkowski functionals 
place limits on the non-linear coupling parameter $f_{\rm NL}$, which
characterizes the amplitude of a quadratic term in
the Bardeen curvature perturbations [Eq.~(\ref{eq:phi})].
It is important to remove the best-fit foreground templates from the 
{\MAP} maps in order to reduce the non-Gaussian Galactic foreground emission.
The cubic statistic measures phase correlations of temperature fluctuations
to find the best estimate of $f_{\rm NL}$ from the foreground-removed, 
weighted average of Q$+$V$+$W maps as $f_{\rm NL}=38\pm 48$ (68\%)
and $-58<f_{\rm NL}<134$ (95\%).
The Minkowski functions measure morphological structures to find 
$f_{\rm NL}=22\pm 81$ (68\%) and $f_{\rm NL}<139$ (95\%), in good 
agreement with the cubic statistic.
These two completely different statistics give consistent results,
validating the robustness of our limits.
Our limits are 20$-$30 times better than the previous ones
\citep{komatsu/etal:2002,santos/etal:2002b,cayon/etal:2002},
and constrain the relative contribution from the non-linear term to the r.m.s.
amplitude of $\Phi$ to be smaller than $2\times 10^{-5}$ (95\%),
much smaller than the limits on systematic errors in the {\MAP} observations.
This validates that the angular power spectrum can fully characterize
statistical properties of the {\MAP} CMB sky maps.
We conclude that the {\MAP} 1-year data do not show 
evidence for significant primordial non-Gaussianity of the form
in equation~(\ref{eq:phi}).
Our limits are consistent with predictions from inflation models
based upon a slowly rolling scalar field, 
$\left|f_{\rm NL}\right|=10^{-2}-10^{-1}$.
The span of all non-Gaussian models, however, is large,
and there are models which cannot be parametrized by
equation~(\ref{eq:phi}) (e.g., \citet{bernardeau/uzan:2002a,bernardeau/uzan:2002b}).
Other forms such as multi-field inflation models and topological defects 
will be tested in the future.

The non-Gaussianity also affects the dark-matter halo mass function $dn/dM$,
since the massive halos at high redshift are sensitive to changes in the
tail of the distribution function of density fluctuations.
Our limits show that the number of clusters that would be newly found
at $z=1$ for $M<10^{15}~M_\odot$ should be within ${}^{+40}_{-10}\%$
of the value predicted from the Gaussian theory.
At higher redshifts, however, much larger effects are still allowed.
The number counts $dN/dz$ at $z=3$ with the limiting mass of 
$3\times 10^{14}~M_\odot$ can be reduced by a factor of 2, or increased 
by more than a factor of 3.
Since the SZ angular power spectrum is primarily sensitive
to massive halos at $z\sim 1$ where the impact of non-Gaussianity is 
constrained to be within 10\%, a measurement of $\sigma_8$ from 
the SZ angular power spectrum is changed by no more than 2\%.
Our results on $dn/dM$ derived in this paper should be taken
as the current {\it observational} limits to non-Gaussian effects
on $dn/dM$.
In other words, this is the uncertainty that we currently have in
$dn/dM$ when the assumption of Gaussian fluctuations is relaxed.

The limits on $f_{\rm NL}$ will improve as the {\MAP} satellite
acquires more data. Monte Carlo simulations show that
the 4-year data will achieve 95\% limit of 80.
This value will further improve with a more proper pixel-weighting
function that becomes the uniform weighting in the signal-dominated
regime (large angular scales) and becomes the $N^{-1}$ weighting 
in the noise-dominated regime (small angular scales).
There is little hope of testing the expected levels of 
$f_{\rm NL}=10^{-2}-10^{-1}$ from simple inflation models,
but some non-standard models can be excluded.

We have detected non-Gaussian signals arising from the residual radio point
sources left unmasked at Q band, characterized by the reduced point-source 
angular bispectrum 
$b_{\rm src}=(9.5\pm 4.4)\times 10^{-5}~{\rm\mu K^3}~{\rm sr}^2$,
which, in turn, gives the point-source angular power spectrum 
$c_{\rm src}=(15\pm 6)\times 10^{-3}~{\rm\mu K^2}~{\rm sr}$. 
This value agrees well with those from the source number counts
\citep{bennett/etal:2003c} and the angular power spectrum analysis
\citep{hinshaw/etal:2003}, giving us confidence on our understanding
of the amplitude of the residual point sources.
Since $b_{\rm src}$ directly measures $c_{\rm src}$ without relying
on extrapolations, any CMB experiments which suffer from the
point-source contamination should use $b_{\rm src}$ to 
quantify $c_{\rm src}$ to obtain an improved estimate of the
CMB angular power spectrum for the cosmological-parameter determinations.

\cite{hinshaw/etal:2003} found that the best-fit power spectrum
to the {\MAP} temperature data has a relatively large $\chi^2$ value, 
corresponding to a chance probability of 3\%.
While still acceptable fit, there may be missing components
in the error propagations over the Fisher matrix.
Since the Fisher matrix is the four-point function of the 
temperature fluctuations, those missing components (e.g., gravitational
lensing effects) may not be apparent in the bispectrum, the three-point 
function. The point-source non-Gaussianity contributes to the Fisher matrix
by only a negligible amount, as it is dominated by the Gaussian 
instrumental noise. 
Non-Gaussianity in the instrumental noise due to the $1/f$ striping 
may have additional contributions to the Fisher matrix; however, since the 
Minkowski functionals, which are sensitive to higher-order moments of 
temperature fluctuations and instrumental noise, do not find significant 
non-Gaussian signals, non-Gaussianity in the instrumental noise is 
constrained to be very small.

\acknowledgements

The {\MAP} mission is made possible by the support of the Office of Space
Sciences at NASA Headquarters and by the hard and capable work of scores of
scientists, engineers, technicians, machinists, data analysts, budget analysts,
managers, administrative staff, and reviewers.
LV is supportd by NASA through Chandra Fellowship PF2-30022 issued by the
Chandra X-ray Observatory center, which is operated by the Smithsonian
Astrophysical Observatory for an on behalf of NASA under contract NAS8-39073.

\appendix
\section{SIMULATING CMB SKY MAPS FROM PRIMORDIAL FLUCTUATIONS}
\label{app:simulations}

In this appendix, we describe how to simulate CMB sky maps from 
generic primordial fluctuations.
As a specific example, we choose to use the primordial Bardeen
curvature perturbations $\Phi({\mathbf x})$, which generate
CMB anisotropy at a given position of the sky 
$\Delta T(\hat{\mathbf n})$ as \citep{komatsu/spergel/wandelt:2003}
\begin{equation}
 \label{eq:basic}
 \Delta T(\hat{\mathbf n})
  =
  T_0 \sum_{lm} Y_{lm}(\hat{\mathbf n}) 
  \int r^2 dr \Phi_{lm}(r) \alpha_l(r),
\end{equation}
where $\Phi_{lm}(r)$ is the harmonic transform of $\Phi({\mathbf x})$ 
at a given comoving distance $r\equiv \left|{\mathbf x}\right|$,
$ \Phi_{lm}(r)\equiv
  \int d^2 \hat{\mathbf n} \Phi(r,\hat{\mathbf n}) 
  Y^*_{lm}(\hat{\mathbf n})$, and
$\alpha_l(r)$ was defined previously (Eq.[\ref{eq:alpha_l}]).
We can instead use isocurvature fluctuations or a mixture of the two.
Equation~(\ref{eq:basic}) suggests that 
$\alpha_l(r)$ is a transfer function projecting 
$\Phi({\mathbf x})$ onto $\Delta T(\hat{\mathbf n})$ through the 
integral over the line of sight.
Since $\alpha_l(r)$ is just a mathematical function, we pre-compute 
and store it for a given cosmology, reducing the 
computational time of a batch of simulations.
We can thus use or extend equation~(\ref{eq:basic}) to compute 
$\Delta T(\hat{\mathbf n})$ for generic primordial fluctuations.

We simulate CMB sky maps using a non-Gaussian model of the form in
equation~(\ref{eq:phi}) as follows.
(1) We generate $\widetilde{\Phi}_{\rm L}(\mathbf k)$ as 
a Gaussian random field in Fourier space for a given initial power 
spectrum $P(k)$, and transform it back to real space to obtain 
$\Phi_{\rm L}({\mathbf x})$.
(2) We transform from Cartesian to spherical coordinates to obtain 
$\Phi_{\rm L}(r,\hat{\mathbf n})$, compute its harmonic coefficients
$\Phi_{lm}(r)$, and obtain a temperature map of the Gaussian part 
$\Delta T_\Phi(\hat{\mathbf n})$ by integrating
equation~(\ref{eq:basic}).
(3) We repeat this procedure for $\Phi_{\rm L}^2({\mathbf x}) - 
V_x^{-1}\int d^3{\mathbf x}\Phi_{\rm L}^2({\mathbf x})$
to obtain a temperature map of the non-Gaussian part
$\Delta T_{\Phi^2}(\hat{\mathbf n})$.
(4) By combining these two temperature maps, we obtain non-Gaussian sky maps 
for any values of $f_{\rm NL}$,
\begin{equation}
 \label{eq:finalmap}
  \Delta T(\hat{\mathbf n}) = 
  \Delta T_\Phi(\hat{\mathbf n}) 
  + f_{\rm NL} \Delta T_{\Phi^2}(\hat{\mathbf n}).
\end{equation}
We do not need to run many simulations individually 
for different values of $f_{\rm NL}$, but run only twice to obtain
$\Delta T_\Phi(\hat{\mathbf n})$ and $\Delta T_{\Phi^2}(\hat{\mathbf
n})$ for a given initial random number seed.
Also, we can combine $\Delta T_\Phi(\hat{\mathbf n})$ for one seed with
$\Delta T_{\Phi^2}(\hat{\mathbf n})$ for the other to make 
realizations for a particular kind of two-field inflation models.
We can apply the same procedure to isocurvature fluctuations with or 
without $\Phi({\mathbf x})$ correlations.

We need the simulation box of the size of the present-day cosmic 
horizon size $L_{\rm box}=2c\tau_0$,
where $\tau_0$ is the present-day conformal time.
For example, $L_{\rm box}\sim 20~h^{-1}~{\rm Gpc}$ is needed for a 
flat universe 
with $\Omega_{\rm m}=0.3$, whereas we need spatial resolution 
of at least $\sim 20~h^{-1}~{\rm Mpc}$ to resolve the last-scattering 
surface accurately. 
From this constraint the number of grid points is at least 
$N_{\rm grid}=1024^3$, and the required amount of physical memory to 
store $\Phi({\mathbf x})$ is at least 4.3~GB.
Moreover, when we simulate a sky map having 786 432 pixels at 
$nside=256$, we need 1.6~GB to store a field in spherical 
coordinate $\Phi(r,\hat{\mathbf n})$, where the number of $r$ evaluated 
for $N_{\rm grid}=1024^3$ is 512.
Since our algorithm for transforming Cartesian into
spherical coordinates requires another 1.6~GB, in total we 
need at least 7.5~GB of physical memory to simulate one sky map.

\begin{figure}
\plotone{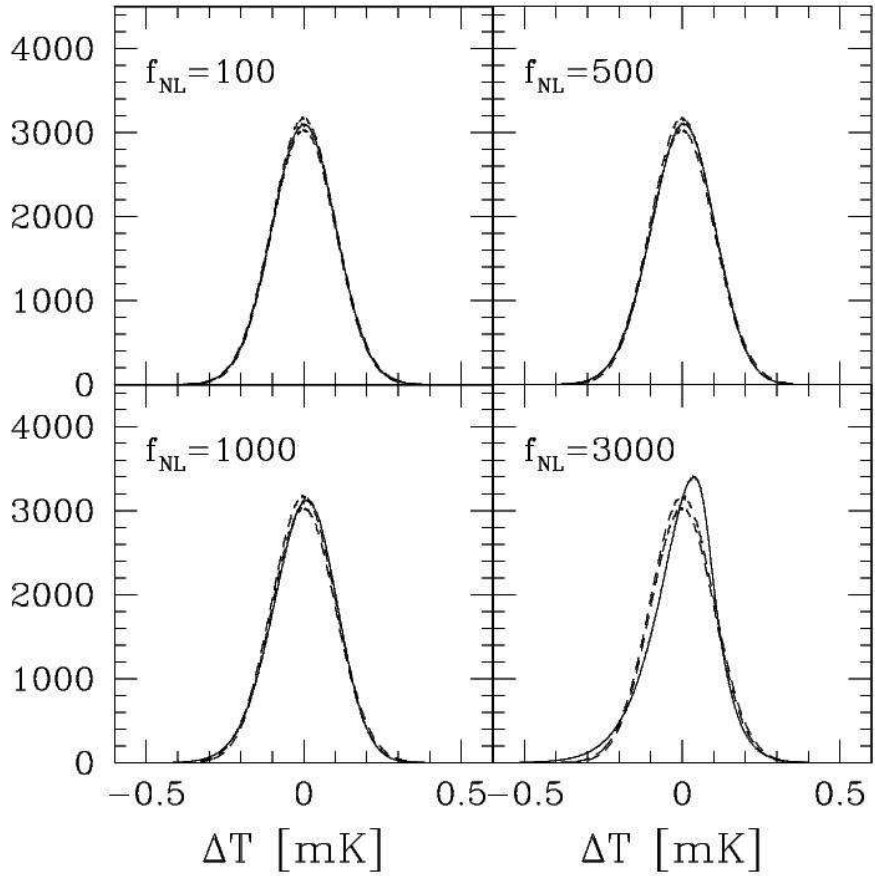}
 \caption{
 One-point PDF of temperature fluctuations measured from simulated 
 non-Gaussian maps (noise and beam smearing are not included). 
 From the top-left to the bottom-right panel the solid lines show 
 the PDF for $f_{\rm NL}=100$, 500, 1000, and 3000, while the dashed 
 lines enclose the r.m.s. scatter of Gaussian realizations
 (i.e., $f_{\rm NL}=0$). \label{fig:pdf}
 }
\end{figure}

We have generated 300 realizations of non-Gaussian sky maps with 
$N_{\rm grid}=1024^3$ and $nside=256$. 
It takes 3 hours on 1 processor of SGI Origin 300 to simulate 
$\Delta T_\Phi(\hat{\mathbf n})$ and $\Delta T_{\Phi^2}(\hat{\mathbf
n})$. We have used 6 processors to simulate 300 maps in one week.
Figure~\ref{fig:pdf} shows the one-point probability density
function (PDF) of temperature
fluctuations measured from simulated non-Gaussian maps (without noise and beam
smearing) compared with the r.m.s. scatter of Gaussian realizations.
We find it difficult for the PDF alone to distinguish 
non-Gaussian maps of $\left|f_{\rm NL}\right|<500$ from Gaussian maps,
whereas the cubic statistic $S_{\rm prim}$ (eq.[\ref{eq:f_NL}]) can easily 
detect $f_{\rm NL}=100$ in the same data sets.

\begin{figure}
\plotone{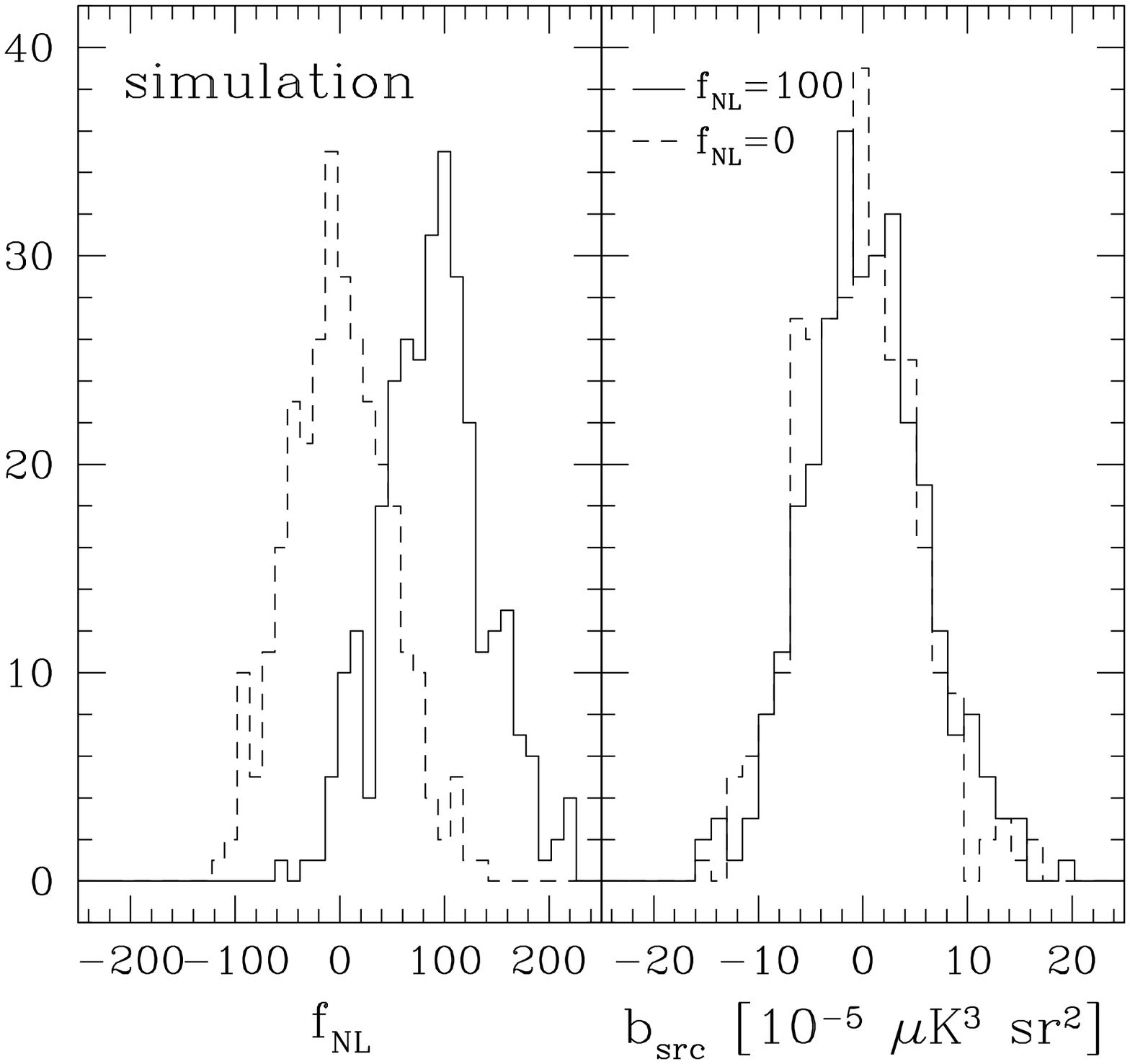}
 \caption{
 The distribution of the non-linear coupling parameter $f_{\rm NL}$ 
 (the left panel) and
 the point-source bispectrum $b_{\rm src}$ (the right panel) measured from
 300 simulated realizations of non-Gaussian maps for
 $f_{\rm NL}=100$ (solid line) and $f_{\rm NL}=0$ (dashed line).
 The simulations include noise properties and window functions of 
 the {\MAP} 1-year data, but do not include point sources. 
 \label{fig:sim_hist}
 }
\end{figure}

We measure $f_{\rm NL}$ on the simulated maps using $S_{\rm prim}$ 
to see if it can accurately recover $f_{\rm NL}$.
Similar tests show the Minkowski functionals to be unbiased and able to 
discriminate different $f_{\rm NL}$ values at levels consistent with the 
quoted uncertainties.
We also measure the point-source angular bispectrum $b_{\rm src}$ to see if 
it returns null values as the simulations do not contain point sources.
We have included noise properties and window functions in the simulations.
Figure~\ref{fig:sim_hist} shows histograms of $f_{\rm NL}$ and $b_{\rm
src}$ measured from 300 simulated maps of $f_{\rm NL}=100$ (solid
lines) and $f_{\rm NL}=0$ (dashed lines).
Our statistics find correct values for $f_{\rm NL}$ and find null values 
for $b_{\rm src}$; thus, our statistics are unbiased, and $f_{\rm NL}$ and
$b_{\rm src}$ are orthogonal to each other as pointed out
by \cite{komatsu/spergel:2001}.

\section{POWER OF THE POINT-SOURCE BISPECTRUM}
\label{app:ptsrc}

In this appendix, we test our estimator for $b_{\rm src}$ and 
$c_{\rm src}$ using simulated Q-band maps of point sources, CMB, and 
detector noise.
The 44~GHz source count model of T98 was
used to generate the source populations. 
The total source count in each realization was fixed to 9043,
the number predicted by T98 to lie between
$S_{\rm min}=0.1\,{\rm Jy}$ and $S_{\rm max}=10\,{\rm Jy}$.
By generating uniform deviates $u\in(0,1)$ and transforming to flux $S$ via
\begin{equation}
u = {{N(>S_{\rm min}) - N(>S)}\over{N(>S_{\rm min}) - N(>S_{\rm max})}},
\end{equation}
we obtain the desired spectrum.
The sources were distributed evenly over the sky and convolved with
a Gaussian profile approximating the Q-band beam.
Flux was converted to peak brightness using the values in Table~8 of 
\citet{page/etal:2003}.
The CMB and noise realizations were not varied between realizations.
The goal in this appendix is to prove that our estimator for 
$b_{\rm src}$ works well and is very powerful in estimating $c_{\rm src}$.

\begin{figure}
\plotone{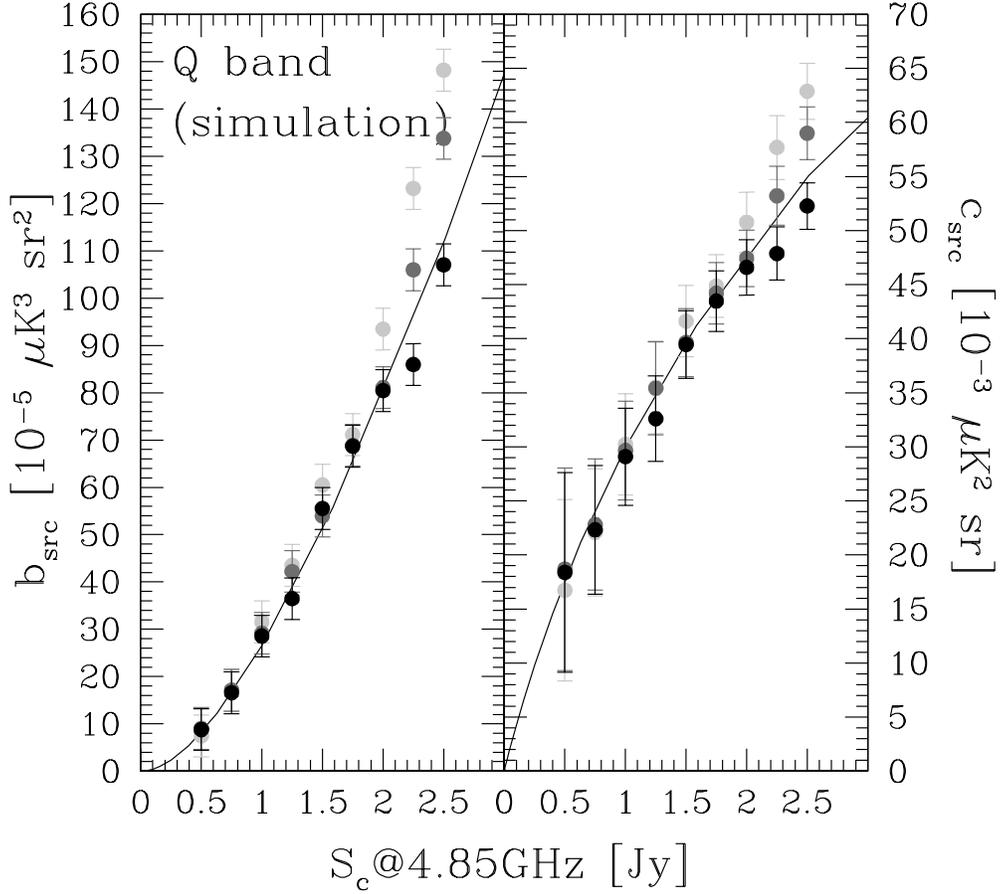}
 \caption{
 Testing the estimator for the reduced point-source bispectrum
 $b_{\rm src}$ [Eq.(\ref{eq:b_ps*})].
 The left panel shows 
 $b_{\rm src}$ measured from a simulated map including point 
 sources and properties of the {\MAP} sky map at Q band,
 as a function of flux cut $S_{\rm c}$ (filled circles).
 Black, dark-gray, and light-gray indicate three different realizations
 of point sources.
 The solid line is the expectation from the 
 input source number counts in the point-source simulation.
 The right panel compares the power spectrum $c_{\rm src}$ estimated
 from $b_{\rm src}$ with the expectation. 
 The error bars are not independent, because the distribution is
 cumulative.
 The behavior for $S_{\rm c}> 2~{\rm Jy}$ shows the cumulative effect 
 of sources with brightness comparable to the instrument noise
 (see text in Appendix~\ref{app:ptsrc}).
 \label{fig:ptsrc_sim}
 }
\end{figure}

The left panel of Figure~\ref{fig:ptsrc_sim} compares the measured 
$b_{\rm src}$ from simulated maps with
the expectations of the simulations.
Black, dark-gray, and light-gray indicate three different realizations
 of point sources.
The measurements agree well with the expectations at $S_{\rm c}<1.75~{\rm Jy}$.
They however show significant scatter at $S_{\rm c}>1.75~{\rm Jy}$, because
our filter for computing $b_{\rm src}$ [eq.~(\ref{eq:filter3})]
does not include contribution from $c_{\rm src}$ to $\tilde{C}_l$, making 
the filter less optimal in the limit of ``too many'' unmasked point sources.
We can see from the figure that $c_{\rm src}$ at $S_{\rm c}>2~{\rm Jy}$
is comparable to or larger than the noise power spectrum for Q band, 
$54\times 10^{-3}~{\mu\rm K}^2~{\rm sr}$.

Fortunately this is not a problem in practice, as we can detect and mask 
those bright sources which contribute significantly to $\tilde{C}_l$.
The residual point sources that we cannot detect (therefore we want to 
quantify using $b_{\rm src}$) should be hidden in the noise having
only a small contribution to $\tilde{C}_l$.
In this faint-source regime $b_{\rm src}$ works well in measuring 
the amplitude of residual point sources, offering a promising way for
estimating $c_{\rm src}$.
The right panel of Figure~\ref{fig:ptsrc_sim} compares $c_{\rm src}$
estimated from $b_{\rm src}$ [Eq.~(\ref{eq:bsrc})] with
the expectations.
The agreement is good for $S_{\rm c}<1.75~{\rm Jy}$, proving that
estimates of $c_{\rm src}$ from $b_{\rm src}$ are unbiased and powerful.
Since $b_{\rm src}$ measures $c_{\rm src}$ directly, we can use 
it for any CMB experiments which suffer from the effect
of residual point sources.
While we have considered the bispectrum only here, the forth order moment
may also be used to increase our sensitivity to the point-source 
non-Gaussianity \citep{pierpaoli:2003}.

\section{THE ANGULAR BISPECTRUM FROM A POTENTIAL STEP}
\label{app:step}

A scalar-field potential $V(\phi)$ with features can generate large 
non-Gaussian fluctuations in CMB by breaking the slow-roll conditions 
at the location of the features
\citep{kofman/etal:1991,wang/kamionkowski:2000}.
We estimate the impact of the features by using a scale-dependent
$f_{\rm NL}$,
\begin{equation}
 \label{eq:SB}
 f_{\rm NL}(\phi)= -\frac5{24\pi G}
 \left(\frac{\partial^2\ln H}{\partial\phi^2}\right),
\end{equation}
which is calculated from a non-linear transformation between the curvature
perturbations in the comoving gauge and the scalar-field fluctuations 
in the spatially flat gauge
\citep{salopek/bond:1990,salopek/bond:1991}.
This expression does not assume the slow-roll conditions.
Although this expression does not include all effects contributing to
$f_{\rm NL}$ during inflation driven by a single field
\citep{maldacena:2002}, we assume that an order-of-magnitude estimate
can still be obtained.

A sharp feature in $V(\phi)$ at $\phi_{\rm f}$ produces
a significantly scale-dependent $f_{\rm NL}(\phi)$ near
$\phi_{\rm f}$ through the derivatives of $H$ in equation~(\ref{eq:SB}).
We illustrate the effects of the steps using the potential features
proposed by \cite{adams/cresswell/easther:2001},
\begin{equation}
 \label{eq:feature}
 V(\phi)= \frac12m_\phi^2\phi^2
 \left[1+c\tanh\left(\frac{\phi-\phi_{\rm f}}{d}\right)\right],
\end{equation}
which has a step in $V(\phi)$ at $\phi_{\rm f}$ with the height $c$
and the slope $d^{-1}$.
\cite{adams/ross/sarkar:1997b} show that the steps are created
by a class of supergravity models in which symmetry-breaking
phase transitions of many fields in flat directions gravitationally 
coupled to $\phi$ continuously generate steps in $V(\phi)$ every 
$10-15$ $e$-folds, giving a chance
for a step to exist within the observable region of $V(\phi)$.

It is instructive to evaluate equation~(\ref{eq:SB}) combined with
equation~(\ref{eq:feature}) in the slow-roll limit, 
$\partial^2\ln H/\partial\phi^2\simeq \frac12\partial^2\ln V/\partial\phi^2$.
For $\left|c\right|\ll 1$, one obtains
\begin{equation}
 \label{eq:SBSR}
 f_{\rm NL}(\phi)
 \simeq
 \frac{5}{24\pi G}
 \left( \frac1{\phi^2} + \frac{c}{d^2}\frac{\tanh x}{\cosh^2x} \right),
\end{equation}
where $x\equiv (\phi-\phi_{\rm f})/d$.
The first term corresponds to a standard, nearly scale-independent 
prediction giving 
$7.4\times 10^{-3}$ at $\phi=3m_{\rm planck}$, 
while the second term reveals a significant scale-dependence.
The function $\tanh x/\cosh^2x$ is a symmetric odd function about $x=0$ 
with extrema of $\pm 0.385$ at $x\simeq \pm 0.66$.
The picture is the following: as $\phi$ rolls down $V(\phi)$ from a 
positive $x>0.66$, $\phi$ gets accelerated at $x\simeq 0.66$, 
reaches constant velocity at $x=0$, decelerates at $x\simeq -0.66$, and 
finally reaches slow roll at $x<-0.66$.
The ratio of the second term in equation~(\ref{eq:SBSR}) to the first at 
the extrema is $\pm 0.385c(\phi/d)^2$.
For example, $c=0.02$ and $\phi/d=300$ (i.e., $d=0.01m_{\rm planck}$) 
make the amplitude of the second term 700 times larger than the first, giving 
$\left|f_{\rm NL}\right|\simeq 5$ at the extrema.
Despite the slow-roll conditions having a tendency to underestimate $f_{\rm NL}$,
it is possible to obtain $\left|f_{\rm NL}\right|>1$. 
Neglecting the first term in equation~(\ref{eq:SBSR}) and converting 
$\phi$ for $k$, one obtains
\begin{equation}
 \label{eq:SBSRk}
 f_{\rm NL}(k)
 \simeq
 \frac{5c}{24\pi G d^2} h_{\rm step}(k)
 \equiv 
 \frac{5c}{24\pi G d^2}\frac{\tanh x_k}{\cosh^2x_k},
\end{equation}
where $x_k\simeq d^{-1}(\partial\phi/\partial\ln k)_{\rm f}(k/k_{\rm f}-1)
= d^{-1}(\dot{\phi}/H)_{\rm f}(k/k_{\rm f}-1)$ for $k-k_{\rm f}\ll k_{\rm f}$.
The slow-roll approximation gives 
$x_k\simeq (4\pi G\phi_{\rm f} d)^{-1}(k/k_{\rm f}-1)$.
Finally, following the method of \cite{komatsu/spergel:2001}, we obtain
the reduced bispectrum of a potential step model, $b^{\rm
step}_{l_1l_2l_3}$,  as
\begin{equation}
 \label{eq:bprim}
 b^{\rm step}_{l_1l_2l_3}
 = 2\left(\frac{5c}{24\pi G d^2}\right)
 \int_0^\infty r^2 dr 
 \left[\beta_{l_1}(r)\beta_{l_2}(r)
 \alpha^{\rm step}_{l_3}(r) + \mbox{(2 permutations)}\right],
\end{equation}
where $\beta_l(r)$ is given by equation~(\ref{eq:beta_l}), and
\begin{equation}
 \label{eq:alpha2}
  \alpha^{\rm step}_l(r) \equiv
  \frac{2}{\pi}\int k^2 dk h_{\rm step}(k) g_{{\rm T}l}(k) j_l(k r).
\end{equation}
The amplitude is thus proportional to $c/d^2$: a bigger (larger $c$) 
and steeper (smaller $d$) step gives a larger bispectrum.
The steepness affects the amplitude more, because the non-Gaussianity
is generated by breaking the slow-roll conditions.

Since $b^{\rm step}_{l_1l_2l_3}$ linearly scales as $c$ for
a fixed $d$, we can fit for $c$ by using exactly the same method as for
the scale-independent $f_{\rm NL}$, but with $\alpha_l(r)$ in 
equation~(\ref{eq:filter1}) replaced by $\alpha^{\rm step}_l(r)$.
The exact form of the fitting parameter in the slow-roll limit is 
$5c/(24\pi G d^2)$.
A reason for the similarity between the two models in methods for
the measurement is explained as follows.
\cite{komatsu/spergel/wandelt:2003} have shown that $B(\hat{\mathbf n},r)$
[Eq.(\ref{eq:filter2})] is a Wiener-filtered, reconstructed map of the 
primordial fluctuations $\Phi(\hat{\mathbf n},r)$.
Our cubic statistic [Eq.(\ref{eq:skewness})] effectively measures the skewness 
of the reconstructed $\Phi$ field, maximizing the sensitivity to the 
primordial non-Gaussianity.
One of the three maps comprising the cubic statistic is however
not $B(\hat{\mathbf n},r)$, but $A(\hat{\mathbf n},r)$
given by equation~(\ref{eq:filter1}). 
This map defines what kind of non-Gaussianity to look for,
or more detailed form of the bispectrum.
For the potential step case, $A_{\rm step}(\hat{\mathbf n},r)$ made of
$\alpha^{\rm step}_l(r)$ picks up the location of the step to measure
$5c/(24\pi G d^2)$  near $k_{\rm f}$, while for the form in 
equation~(\ref{eq:phi}),
$A(\hat{\mathbf n},r)$ explores all scales on equal footing to measure the
scale-independent $f_{\rm NL}$.
 
The distinct features in $k$ space are often smeared out in $l$ space
via the projection.
This effect is estimated from equation~(\ref{eq:alpha2}) as follows.
The function $h_{\rm step}(k)$ near $k_{\rm f}$ is accurately approximated 
by $h_{\rm step}(k)\simeq 0.385 \sin(2x_k)$, which
has a period of $\Delta k=4\pi^2G\phi_{\rm f} dk_{\rm f}$.
On the other hand, the radiation transfer function $g_{{\rm T}l}(k)$ behaves as
$j_l(kr_*)$ where $r_*$ is the comoving distance to the photon decoupling
epoch, and $g_{{\rm T}l}(k) j_l(k r)$ behaves as $j_l^2(kr_*)$
(the integral is very small when $r\neq r_*$).
The oscillation period of this part is thus $\Delta k=\pi/r_*$ for
$kr_*>l$.
A ratio of the period of $h_{\rm step}(k)$ to that of 
$g_{{\rm T}l}(k) j_l(k r)$ is then estimated as 
$4\pi G\phi_{\rm f}dr_*k_{\rm f}\simeq
(l_{\rm f}/3)(d/0.01m_{\rm planck})(\phi_{\rm f}/3m_{\rm planck})$, where
$l_{\rm f}\equiv k_{\rm f}r_*$ is the angular wave number of the location
of the step.
We thus find that $h_{\rm step}(k)$ oscillates much more slowly than 
the rest of the integrand in equation~(\ref{eq:alpha2}) for 
$l_{\rm f}\gg 1$.

What does it mean? 
It means that the results would look as if there were two distinct
regions in $l$ space where $f_{\rm NL}$ is very large:
a positive $f_{\rm NL}$ is found at $l<l_{\rm f}$ and
a negative one at $l>l_{\rm f}$. 
The estimated location is $l/l_{\rm f}\simeq 1\pm 0.66(4\pi G\phi_{\rm f}d)
\simeq 1\pm 0.2(d/0.01m_{\rm planck})(\phi_{\rm f}/3m_{\rm planck})$;
thus, the positive and negative regions are separated in $l$ by only
40\%, making the detection difficult when many $l$ modes are combined
to improve the signal-to-noise ratio.
The two extrema would cancel out to give only small signals.
In other words, it is still possible that non-Gaussianity from
a potential step is ``hidden'' in our measurements shown in 
Figure~\ref{fig:f_NL}.
Note that the cancellation occurs because of the point symmetry of 
$h_{\rm step}(k)$ about $k=k_{\rm f}$.
If the function has a knee instead of a step, then the cancellation
does not occur and there would be a single region in $l$ space where 
$\left|f_{\rm NL}\right|$ is large \citep{wang/kamionkowski:2000}.
Note that our estimate in this Appendix was based upon 
equation~(\ref{eq:SBSR}), which uses the slow-roll approximations.
While instructive, since the slow-roll approximations break down near 
the features, our estimate may not be very accrate.
One needs to integrate the equation of motion of the scalar field 
to evaluate equation~(\ref{eq:SB}) for more accurate estimations of 
the effect.

\end{document}